\documentclass[11pt,twoside]{article}
\usepackage{graphicx,floatflt,amssymb,epsfig,epsf,pstricks}
\usepackage{times}
\usepackage{a4wide}
\usepackage{cite}
\usepackage{slashed}

  \usepackage{fancyheadings}
  \advance \headheight by 3.0truept       

\newlength{\nseparation}
\newenvironment{nfigure}[1]
        {\begin{figure}[#1]\hrule\vspace{\nseparation}\par}
        {\vspace{\nseparation}\par \hrule \end{figure}}

\def\bea {\begin{eqnarray}}
\def\eea {\end{eqnarray}}

\newcommand{\dd}{\ensuremath{D\!-\!\Dbar\,}}
\newcommand{\ddm}{\dd\ mixing}

\newcommand{\eq}[1]{Eq.~(\ref{#1})}
\newcommand{\eqsand}[2]{Eqs.~(\ref{#1}) and (\ref{#2})}

\newcommand{\Dbar}{\,\overline{\!D}}

\newcommand{\no}{\nonumber}
\newcommand{\nn}{\nonumber \\}
\newcommand{\ov}[1]{\overline{#1}}
\newcommand{\lt}{\left}
\newcommand{\rt}{\right}
\newcommand{\ps}{p\hspace{-0.44em}/\hspace{0.06em}}
\newcommand{\gev}{\mbox{GeV}}
\newcommand{\tev}{\mbox{TeV}}
\newcommand{\fig}[1]{Fig.~\ref{#1}}
\newcommand{\ds}{\displaystyle}

\newcommand{\real}{\mathrm{Re}\,}

\begin{document}
\thispagestyle{plain}
\parbox[t]{0.4\textwidth}{TTP08--41}
\hfill October 2008 \\

\begin{center}
\boldmath
{\Large \bf Supersymmetric renormalisation of the CKM matrix 
            and\\[2mm] new constraints on the squark mass matrices  }\\
\unboldmath
\vspace*{1cm}
\renewcommand{\thefootnote}{\fnsymbol{footnote}}
Andreas Crivellin and Ulrich Nierste \\
\vspace{10pt}
{\small
              {\em Institut f\"ur Theoretische Teilchenphysik\\
               Karlsruhe Institute of Technology, 
               Universit\"at Karlsruhe,\\ D-76128 Karlsruhe, Germany}} \\
\normalsize
\end{center}

\begin{abstract}
  We compute the finite renormalisation of the Cabibbo-Kobayashi-Maskawa (CKM)
  matrix induced by gluino--squark diagrams in the MSSM with non-minimal
  sources of flavour violation. Subsequently we derive bounds on the
  flavour--off--diagonal elements 
  of the squark mass matrices by requiring that the radiative corrections to
  the CKM elements do not exceed the experimental values.  Our constraints on
  the associated dimensionless quantities $\delta^{d\,LR}_{ij}$, $j>i$, are
  stronger than the bounds from flavour-changing neutral current (FCNC)
  processes if gluino and squarks are heavier than 500$\,\gev$. Our bound on
  $|\delta^{u\,LR}_{12}|$ is stronger than the FCNC bound from \ddm\ for
  superpartner masses above 900$\,\gev$.  We further find a useful bound on
  $|\delta_{13}^{u\,LR}|$, for which no FCNC constraint is known.   Our
  results imply that it is still possible to generate all observed flavour
  violation from the soft supersymmetry-breaking terms without conflicting
  with present-day data on FCNC processes.  We suggest that a flavour symmetry
  renders the Yukawa sector flavour--diagonal and the trilinear
  supersymmetry-breaking terms are the spurion fields breaking this flavour
  symmetry. We further derive the dominant supersymmetric radiative
  corrections to the couplings of charged Higgs bosons and charginos to quarks
  and squarks.
\end{abstract}

\section{Introduction}
The generic Minimal Supersymmetric Standard Model (MSSM) contains a
plethora of new sources of flavour violation, which reside in the
supersymmetry--breaking sector. The origin of these flavour--violating
terms is easily understood: In the Standard Model (SM) the Yukawa
matrices are diagonalised by unitary rotations in flavour space and the
resulting basis defines the quark mass eigenstates. If the same
rotations are carried out on the squark fields of the MSSM, one obtains
the super--CKM basis in which no tree--level FCNC couplings are present.
However, neither the $3\times 3$ mass terms ${\bf M}^2_{\tilde q}$,
${\bf M}^2_{\tilde d}$ and ${\bf M}^2_{\tilde u}$ of the left--handed
and right--handed squarks nor the trilinear Higgs--squark--squark
couplings are necessarily diagonal in this basis. The trilinear
$\ov{Q}H_d {\bf A}^d d_R $ and $\ov{Q}H_u {\bf A}^u u_R $ terms induce
mixing between left--handed and right--handed squarks after the Higgs
doublets $H_d$ and $H_u $ acquire their vacuum expectation values (vevs)
$v_d$ and $v_u$, respectively. In the conventions of
Ref.~\cite{Gabbiani:1996} the full $6\times 6$ mass matrix for the
down--squarks reads

\begin{equation}
\renewcommand{\arraystretch}{1.4}
 M_{\tilde d}^2  = \left( {\begin{array}{*{20}c}
    {\left(M_{1L}^{\tilde d}\right)^2} & {\Delta _{12}^{\tilde{d}\,LL} } & {\Delta
    _{13}^{\tilde{d}\,LL} } & {\Delta _{11}^{\tilde{d}\,LR} } & {\Delta
    _{12}^{\tilde{d}\,LR} } & {\Delta _{13}^{\tilde{d}\,LR} }  \\ 
    {{\Delta _{12}^{\tilde{d}\,LL}}^* } & {\left(M_{2L}^{\tilde d}\right)^2 } & {\Delta
    _{23}^{\tilde{d}\,LL} } & {{\Delta _{12}^{\tilde{d}\,RL}}^* } & {\Delta
    _{22}^{\tilde{d}\,LR} } & {\Delta _{23}^{\tilde{d}\,LR} }  \\ 
    {{\Delta _{13}^{\tilde{d}\,LL}}^* } & {{\Delta _{23}^{\tilde{d}\,LL} }^*} &
    {\left(M_{3L}^{\tilde d}\right)^2 } & {{\Delta _{13}^{\tilde{d}RL}}^* } & {\Delta
    _{23}^{RL*} } & {\Delta _{33}^{\tilde{d}\,LR} }  \\ 
    {{\Delta _{11}^{\tilde{d}\,LR}}^* } & {\Delta _{12}^{\tilde{d}RL} } & {\Delta
    _{13}^{\tilde{d}RL} } & {\left(M_{1R}^{\tilde d}\right)^2 } & {\Delta
    _{12}^{\tilde{d}\,RR} } & {\Delta _{13}^{\tilde{d}\,RR} }  \\ 
    {{\Delta _{12}^{\tilde{d}\,LR}}^* } & {\Delta _{22}^{\tilde{d}\,LR*} } & {\Delta
    _{23}^{\tilde{d}RL} } & {{\Delta _{12}^{\tilde{d}\,RR}}^* } & {\left(M_{2R}^{\tilde
    d}\right)^2 } & {\Delta _{23}^{\tilde{d}\,RR} }  \\ 
    {{\Delta _{13}^{\tilde{d}\,LR}}^* } & {{\Delta _{23}^{\tilde{d}\,LR}}^* } & {{\Delta
    _{33}^{\tilde{d}\,LR}}^* } & {{\Delta _{13}^{\tilde{d}\,RR}}^* } & {{\Delta
    _{23}^{\tilde{d}\,RR}}^* } & {\left(M_{3R}^{\tilde d}\right)^2 }  \\ 
 \end{array}} \right)\label{massmatrix}
\end{equation}

and the up--type squark mass matrix is defined in an analogous way with
$\tilde d$ replaced by $\tilde u$. Here the $\Delta _{ij}^{\tilde q\,LR}$,
$i,j=1,\ldots 3$, are related to the trilinear terms as
\begin{eqnarray}
\Delta _{ij}^{\tilde d \,LR} &=& {A}^{d}_{ij} v_d \;=\; 
{A}^{d}_{ij} v \cos \beta, 
\quad
\Delta _{ij}^{\tilde u \, LR} \;=\; 
{A}^{u}_{ij} v_u \;=\; {A}^{u}_{ij} v \sin \beta 
 \qquad \textrm{for} \: j>i.~~
\label{dea}
\end{eqnarray}

We normalise the Higgs vevs as $v=\sqrt{v_u^2+v_d^2}\simeq 174\,\gev$ and
define $\tan\beta =v_u/v_d$ as usual. The complete squark mass matrix is given
in the appendix, where we also elaborate on the relationship between weak
bases and the super--CKM basis.  The diagonalisation of $M_{\tilde q}^2$
involves a rotation of the squark fields in flavour space which leads to
various flavour--changing neutral couplings. In particular the gluino now
couples to quarks and squarks of different generations and FCNC processes
occur through strong gluino--squark loops, which easily dominate over the
highly CKM--suppressed weak loops of the SM. Anticipating the smallness of the
off--diagonal elements in $M_{\tilde{d},\tilde{u}}^2$ one can alternatively
work in the super--CKM basis and treat the $\Delta _{ij}^{\tilde q\,XY}$'s
(with $X,Y=L$ or $R$) as perturbations
\cite{Hall:1985dx,Gabbiani:1996,Misiak:1997ei,Buras:1997ij}. It is customary
to define the dimensionless quantities

\begin{equation}
\delta^{q \,XY} _{ij}  = 
  \frac{\Delta^{\tilde q\, XY}_{ij}}{\frac{1}{6}\sum\limits_s 
     {\left[M_{\tilde q}^2\right]_{ss}}} .\label{defde}
\end{equation}

In the current era of precision flavour physics stringent bounds on these
parameters have been derived from FCNC processes, by requiring that the
gluino--squark loops do not exceed the measured values of the considered
observables \cite{Hagelin:1992, Gabbiani:1996, Ciuchini:1998ix, Borzumati:1999,
  Becirevic:2001,Silvestrini:2007,Ciuchini:2007cw}. In the recent analysis of
Ref.~\cite{Silvestrini:2007} the strongest constraint has been obtained on
$\delta^{\tilde d \,LR}_{23}$ with $\left|\delta^{\tilde d \,LR}_{23}\right| <
10^{-3}$ for 
$\sqrt{\left[M_{\tilde q}^2\right]_{ss}}=m_{\tilde{g}}=350\, \gev$.

In this paper we show that charged--current processes give competitive bounds
on the $\delta^{\tilde q \,LR}_{ij}$'s. This is surprising, because here a
supersymmetric loop competes with a SM tree--level coupling. However, the
flavour structure of the SM is governed by very small Yukawa couplings: In a
weak basis with a diagonal up--type Yukawa matrix $Y^u$ the off--diagonal
elements of the down--type Yukawa matrix $Y^d$ range from $|Y^d_{31}|\sim
10^{-7}$ to $|Y^d_{23}|\sim 6\cdot 10^{-4}$ at the relevant scale of $M_{\rm
  SUSY} = {\cal O}(M^{\tilde q}_s,m_{\tilde{g}})$.  The impact of
supersymmetric loop corrections to these couplings is most easily understood
in the decoupling limit $M_{\rm SUSY} \gg v$: The tree--level and
loop--induced Higgs couplings to down--type quarks are shown in
Fig.~\ref{fig:dec}. After electroweak symmetry breaking all diagrams
contribute to the quark mass matrix.  The loop--induced contributions are
comparable in size to the tree--level term $Y^q_{fi} v_q$ if roughly $|
  A^q_{fi}|/M_{\rm SUSY}\approx 16\pi^2 |Y^q_{fi}| $ or (for down--type
quarks) $|Y^d_{fj} \Delta^{\tilde d \,LL}_{ji} \mu| 
\tan\beta/M_{\rm SUSY}^3 \approx 16\pi^2 |Y^d_{fi}| $, respectively.
(Here $\mu$ is the Higgsino mass parameter.) 

\begin{nfigure}{t}
\includegraphics[width=1\textwidth]{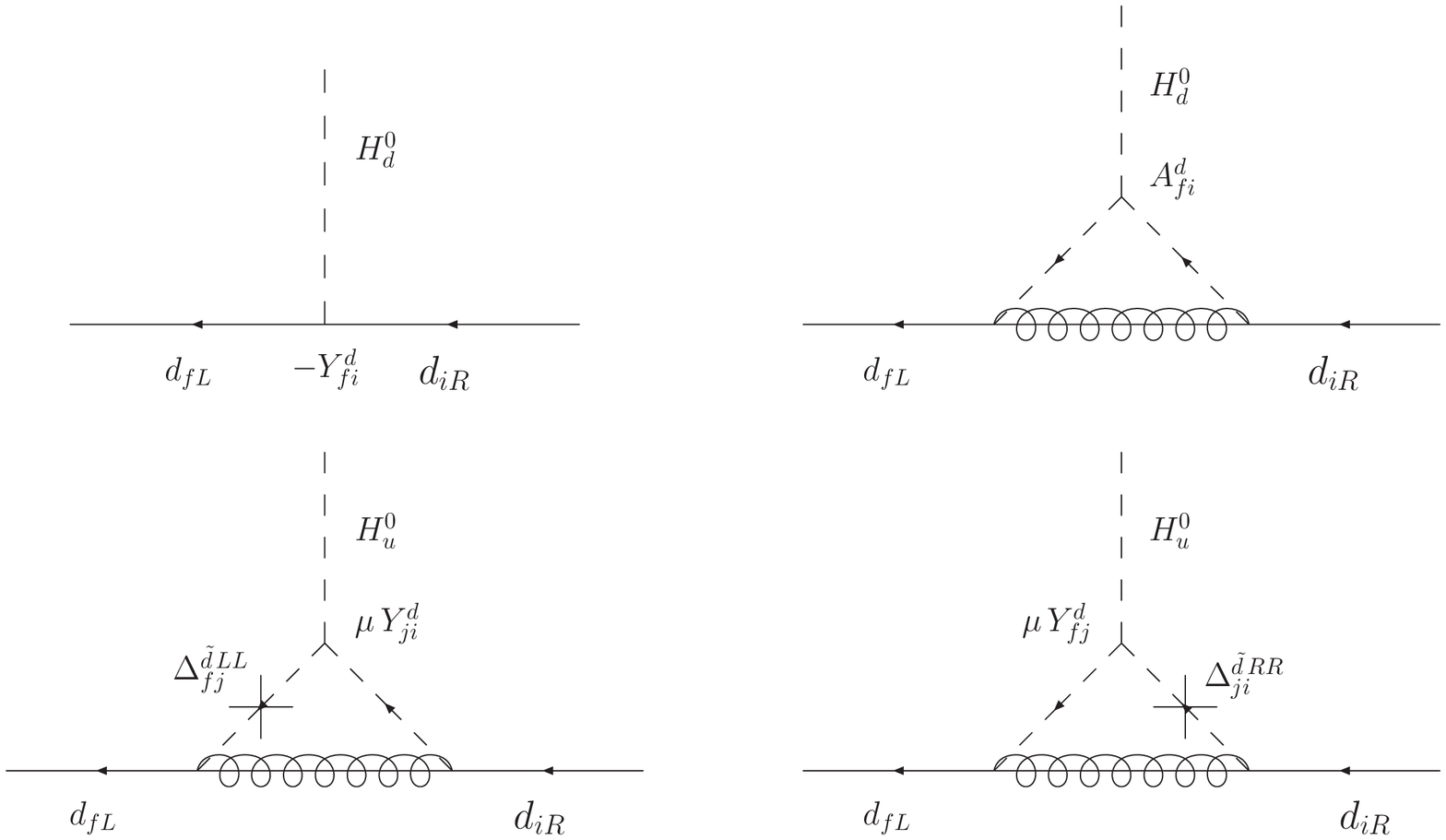}
\caption{Tree--level coupling with $Y^d_{ij}$ and FCNC loop 
  corrections with $A^d_{fi}$ (upper row) and 
  $\Delta _{fi}^{\tilde{d}\,LL,RR}$
  (lower row) in the mass insertion approximation for $M_{\rm SUSY} \gg v$.
  Replacing the Higgs fields by their vevs gives the contributions to the
  down--type quark mass matrix.  The lower diagrams contribute to the mass
  matrix with an enhancement factor of $\tan\beta=v_u/v_d$ compared to the
  other two contributions.}\label{fig:dec}
\end{nfigure}

In Ref.~\cite{Hall:1993gn} it was pointed out that such
corrections constitute an important modification of the relation between
the mass and the Yukawa coupling of the bottom quark.  Subsequent papers
studied the analogous corrections to the whole down-quark mass matrix
for the case of minimal flavour violation (MFV), i.e.\ diagonal matrices
$M_{\tilde u}^2$ and $M_{\tilde d}^2$ in the super-CKM basis
\cite{Hamzaoui:1998nu,Babu:1999hn,Isidori:2002qe,Buras:2002vd}. Here the
key effect of the supersymmetric loop correction is the generation of
effective FCNC couplings of neutral Higgs bosons.  One can proceed along
these lines to calculate the shift in the CKM elements induced by
squark-gluino loops: The quark mass matrix calculated from the diagrams
in \fig{fig:dec} is diagonalised in the usual way yielding the loop-corrected 
CKM matrix $V$. This has been done for the MFV case in
Ref.~\cite{Blazek:1995nv} and for the generic case in
Ref.~\cite{Ferrandis:2004}. As a disadvantage, this method is only valid
in the decoupling limit $M_{\rm SUSY} \gg v$. In particular, this is a
questionable approximation for the top quark, whose mass must be set to
zero in the diagrams of \fig{fig:dec}.  Another difficulty is the
appearance of the Yukawa matrices in the result of these diagrams, while
we only have experimental information on the CKM matrix and the quark
masses. To calculate the Yukawa matrices from the latter, one has to
invert the relations between the loop-corrected mass matrices and the
Yukawa couplings. But in a phenomenological application it is desirable
to have the loop-corrected $V$ directly expressed in terms of CKM
elements and (s)particle masses. All these drawbacks can be avoided if
one renormalises the CKM matrix directly, as it has been done within the
Standard Model in Ref.~\cite{Denner:1990}. This method further involves
only physical quantities and thereby bypasses another pitfall of the
afore-mentioned calculation from the diagrams in \fig{fig:dec}: For
instance, one might be tempted to derive strong bounds on
$\delta^{\tilde d\,RR}_{ij}$ from the lower right diagram of
\fig{fig:dec}. But the result of this diagram can be absorbed into an
unphysical rotation in flavour space of the right-handed quarks and no
such bounds can be found.
  
The main results of our paper are new stringent bounds on the
flavour-off-diagonal entries $\delta^{\tilde q \,LR}_{ij}$ of the squark
mass matrices. These bounds are derived from a fine-tuning argument, by
requiring that no large numerical cancellations should occur between the
tree-level CKM elements and the supersymmetric loop corrections.
Translated to fundamental parameters in the lagrangian, this means that
the loop diagrams in \fig{fig:dec} involving the trilinear
supersymmetry-breaking terms shall not exceed the values of the
tree-level Yukawa couplings. This reasoning is modeled after the
standard line of arguments used to justify low-scale supersymmetry:
Large cancellations between the bare Higgs boson masses and loop
corrections must be avoided, leading to superparticle masses at or below
the \tev\ scale. This argument involves two unphysical quantities, the
bare mass and the corresponding radiative corrections.  In our case the
quantities $Y^q_{ij}$ and $A^q_{ij}$ are separately unobservable as long
as only low-energy quantities are studied.  However, once Higgs or
chargino couplings to squarks are studied, different combinations of
$Y^q_{ij}$ and $A^q_{ij}$ can be investigated and our assumption about
the absence of fine-tuned cancellations can be tested in principle as
discussed in Sect.~\ref{sect:4}. Another viewpoint on the subject is
provided by 't~Hooft's naturalness criterion, which links the smallness
of a quantity to a symmetry which is broken by a small parameter. The
rough size of the symmetry-spoiling parameter can be infered from the
size of the studied quantity.  In the case of the small elements of the
Yukawa matrices the protecting symmetry is a flavour symmetry, which
corresponds to independent rotations of left-handed and right-handed
fermion fields in flavour space. In the SM the only parameters breaking
this symmetry are the small Yukawa couplings. In the generic MSSM the
flavour symmetries are broken by both the Yukawa couplings and the soft
supersymmetry-breaking terms and the natural way to restore the
protecting symmetry is to set the small parameters in both sectors to
zero.  Scenarios in which the $\delta^{\tilde q \,LR}_{ij}$'s
substantially exceed the bounds derived in this paper are therefore
unnatural in 't Hooft's sense.

Our paper is organised as follows: In Sect.~\ref{sect:2} we calculate the 
one-loop renormalisation of the CKM matrix by supersymmetric QCD effects.
In Sect.~\ref{sect:3} we use our results to derive constraints on the 
elements $\Delta _{ij}^{\tilde q\,LR}$ and $\Delta _{ij}^{\tilde q\,LL}$
of the squark mass matrices in \eq{massmatrix}. Here we also reappraise the
idea that flavour violation solely originates from supersymmetry breaking. 
In Sect.~\ref{sect:4} we apply our results to the renormalisation of 
charged-Higgs and chargino couplings to quarks and squarks. 
Finally we conclude. Conventions and Feynman rules are collected in an
appendix.

\section{Renormalisation of the CKM matrix}\label{sect:2}%
To calculate the desired renormalisation of the CKM matrix we must
consider squark-gluino loop corrections to the coupling of the $W$ boson
to quarks.  There are two possible contributions, the self-energy
diagrams of \fig{fig:W} and the proper vertex correction. In the limit
$M_{\rm SUSY} \gg v$ the self-energy contributions reproduce the results
of the diagrams in \fig{fig:dec}. (For a discussion of this feature in
the MFV case see Refs.~\cite{Hall:1993gn,Carena:1999}.)  From the
considerations in the Introduction we know that we need some parametric
enhancement (by e.g.\ a factor of $|A^q_{fi}|/(M_{\rm SUSY}
|Y^q_{fi}|)\gg 1$) to compensate the loop suppression and the diagrams of
\fig{fig:W} involve such enhancement factors.  The vertex diagrams
involving a $W$ coupling to squarks are not enhanced and moreover suffer
from gauge cancellations with non-enhanced pieces from the
self-energies. Therefore we only need to consider self-energies, just as
in the case of the electroweak renormalisation of $V$ in the SM
\cite{Denner:1990}.
\begin{nfigure}{t}
\includegraphics[width=1\textwidth]{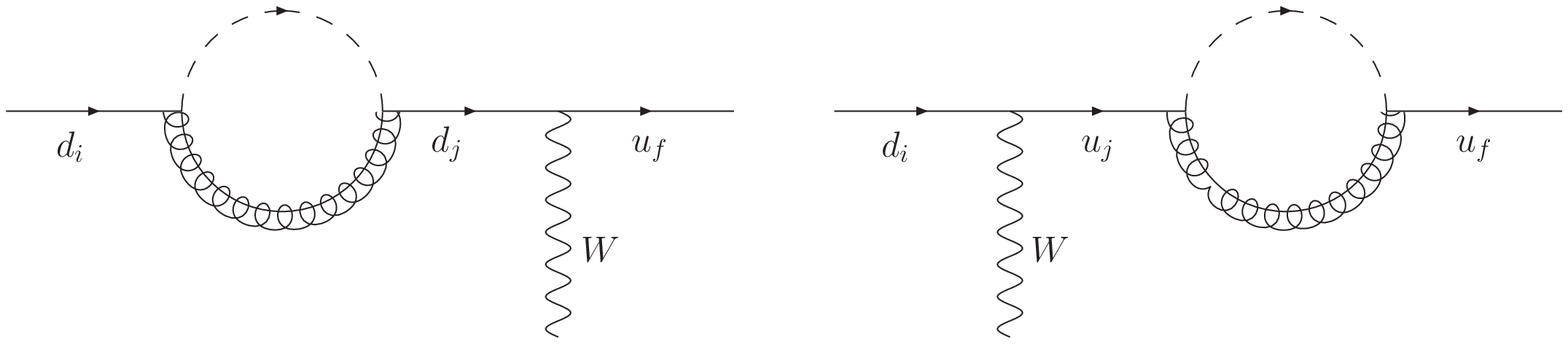}
\caption{One-loop corrections to the CKM matrix from the down and 
  up sectors. We denote the results of the left and right diagrams by 
  $D_{L\, fi}$ and $D_{R\, fi}$, respectively. \label{fig:W}}
\end{nfigure}

From now on we work in the super-CKM basis unless stated otherwise.
Since we work beyond tree-level, we have to clarify how we define the
super-CKM basis in the presence of radiative corrections: Starting from
some weak basis with Yukawa matrices $\mathbf{Y^d}$ and $\mathbf{Y^u}$
we perform the usual rotations in flavour space
\begin{eqnarray} 
d_{L,R} & \to & U_{L,R}^{(0)\,d} d_{L,R}, \qquad\qquad 
u_{L,R} \; \to \; U_{L,R}^{(0)\,u} u_{L,R}  \label{rotq}
\end{eqnarray} 
to diagonalise $\mathbf{Y^q}$ and the tree-level mass matrices
$\mathbf{m_q^{(0)}}=\mathbf{Y^q} v_q$ and apply the same rotations to
$\widetilde d_{L,R}$ and $\widetilde u_{L,R}$.  This defines the super-CKM
basis in which the elements of $M_{\tilde q}^2$ in \eq{massmatrix} are
defined.  The tree-level CKM matrix is then given by
\begin{equation}
V^{(0)} = U_L^{(0)\,u\dag} U_L^{(0)\,d} \label{defckm0} 
\end{equation}
To fix the relation between $V^{(0)}$ and the physical CKM matrix $V$ we
must define a renormalisation scheme.  First note that all radiative
corrections discussed in this paper are finite, so that the notion of
minimal renormalisation means that all counter-terms are simply equal to
zero. Two possibilities come to mind:
\begin{itemize}
\item[i)] Minimal renormalisation of $V$: The Lagrangian contains  
          diagonal Yukawa matrices and $V^{(0)}$ without counter-terms, 
          while the measured CKM matrix $V$ differs from $V^{(0)}$
          by the radiative corrections in \fig{fig:W}. Recall that for 
          $m_j\neq m_i$ one can treat the diagrams of \fig{fig:W} in the
          same way as genuine vertex corrections, i.e.\ there
          is no need to truncate such diagrams or to introduce
          matrix-valued wave function renormalisations \cite{ln}.             
\item[ii)] On-shell renormalisation of  $V$: The Lagrangian contains 
          finite counter-terms to cancel the flavour-changing
          self-energies of \fig{fig:W}. 
          These counter-terms arise from a perturbative   
          unitary rotation of the quark fields in flavour space,              
          $q_{L,R}  \to  [1+\delta U_{L,R}^q] q_{L,R}$
          \cite{Denner:1990}. This in turn induces a counter-term 
          \begin{equation}          
            \delta V = \delta U_L^{u\dag} V^{(0)} + V^{(0)} \delta U_L^d 
          \label{defdu}
          \end{equation}
          to the CKM matrix. In the on-shell scheme we can identify
          $V=V^{(0)}$, but after the extra rotation of the quark fields
          we are no more in the super-CKM basis and the bare Yukawa
          matrices $Y^d$ and $Y^u$ are no more diagonal.\footnote{That
            is, in our Feynman diagrammatic approach the FCNC Higgs
            couplings of
           Refs.~\cite{Hamzaoui:1998nu,Babu:1999hn,Isidori:2002qe,Buras:2002vd}
            enter the Lagrangian through a finite FCNC counter-term to
            Yukawa couplings.}
\end{itemize}

We choose method i), because it involves the super-CKM basis, so that we can
immediately use the $\Delta^{\tilde q\, XY}_{ij}$'s defined in
\eq{massmatrix}, permitting a direct comparison with FCNC analyses.  This
issue of the definition of $\Delta^{\tilde q\, XY}_{ij}$ formally goes beyond
the one-loop order, but is numerically highly relevant, because the tree-level
elements $V^{(0)}_{ij}$ and the finite counter-terms $[\delta U^q_L]_{ij}$ are
similar in size: If one works in an alternative basis in which the (s)quark
superfields are rotated by $[1+\delta U_{L,R}^q]$, the off-diagonal elements
$\Delta^{\tilde q\, XY}_{ij}$ of the squark mass matrices can substantially
differ from those of our definition of the super-CKM basis.

We also need to address the renormalisation scheme used for the quark masses:
The supersymmetric loops are subtracted on-shell, so that we can use the
masses which are extracted from measurements using SM formulae.  While we do
not consider gluonic QCD corrections in this paper, we assume that an $\ov{\rm
  MS}$ prescription is used for the latter. That is, we take $\ov{\rm MS}$
values for the quark masses in our numerical analyses.  This procedure is
guided by the decoupling limit discussed in the Introduction: The Yukawa
couplings of \fig{fig:dec} enter these diagrams as short-distance quantities
defined in a mass-independent scheme such as $\ov{\rm MS}$ and are evaluated
at a scale of order $M_{\rm SUSY}$.  The effective couplings are then evolved
down to a low scale at which the quark masses and $V$ are calculated, yielding
quark masses in the $\ov{\rm MS}$ scheme, yet with decoupled 
(i.e.\ on-shell subtracted) SUSY loops.

The self-energies can be divided into a chirality-flipping and a
chirality-conserving part ($q=u,d$ and $i,f=1,2,3$ labels the incoming and
outgoing quark flavours, respectively):

\begin{equation}
\Sigma _{fi}^q (p) = 
\Sigma^{q\,RL}_{fi} \left( {p^2 } \right) P_L  + 
\Sigma^{q\,LR}_{fi} \left( {p^2 } \right) P_R  + 
\ps  \left[ \Sigma^{q\,LL}_{fi} \left( {p^2 } \right) P_L  + 
            \Sigma^{q\,RR}_{fi} \left( {p^2 } \right) P_R \right]
\label{Zerlegung}
\end{equation}

Since the SUSY particles are much heavier than the five lightest quarks,
it is possible to expand in the external momentum, unless one external
quark is the top.  In the following we consider the self-energies with
only light external quarks and return to the case with a top
quark at the end of this section.

We now write the result of the left diagram of \fig{fig:W}
(omitting external spinors) as

\begin{displaymath}
i\frac{g_2}{\sqrt{2}}\gamma^{\mu} P_L \, D_{L\, fi}
\end{displaymath}

with

\begin{equation}
  D_{L\, fi} =  \sum_{j\neq i} V_{fj}
    \frac{m_{d_j} \left(\Sigma^{d\,RL}_{ji} + m_{d_i} \Sigma^{d\,RR}_{ji}
    \right) \, +\, 
    m_{d_i} \left(\Sigma^{d\,LR}_{ji} + m_{d_i} \Sigma^{d\,LL}_{ji}
    \right)
  }{m_{d_i}^2-m_{d_j}^2}. \label{selfd}
\end{equation}

The diagram with $j=i$ is treated as in the case without flavour mixing,
i.e.\ the self-energy is truncated and contributes to the LSZ factor
in the usual way.  The right diagram $D_{R\, fi}$ involving $\Sigma
_{fj}^u$ instead is obtained in a similar way. Since the quarks
are light compared to the heavy SUSY particles, we can evaluate the
self-energies in \eq{selfd} at $p^2=0$.  $\Sigma^{q\,RL}_{fi}$ and
$\Sigma^{q\,LR}_{fi}$ have mass dimension 1, while $\Sigma^{q\,LL}_{fi}$
and $\Sigma^{q\,RR}_{fi}$ are dimensionless.  The chirality-flipping
self-energies involve at least one power of $v$ and at least one factor
of a trilinear term or a Yukawa coupling multiplied by $\mu$. To first
order in flavour-changing SUSY parameters these factors contribute to
$D_L$ with a parametric enhancement of $|A^q_{ji}| v_q/(M_{\rm SUSY}
\max (m_{q_i},m_{q_j}))$ or (using $m_{q_i}^{(0)}=Y^q_{ii} v_q$) of $|
\Delta^{\tilde d\,LL}_{ji} \mu| \tan\beta/M_{\rm SUSY}^3 $. Thus we find
the enhancement factors which we inferred earlier from \fig{fig:dec}.
Clearly, the terms with $\Sigma^{d\,LL}_{fi}$ and $\Sigma^{d\,RR}_{fi}$
in \eq{selfd} are suppressed by $m_{d_{i,j}}/M_{\rm SUSY}$ compared to
the chirality-flipping contributions and are therefore negligible.

The $LR$ and $RL$ self-energies are 
\begin{equation}
\Sigma^{q\,RL,LR}_{fi} (p^2=0) =
\frac{{2m_{\tilde g} }}{{3\pi }}\alpha _s (M_{\rm SUSY})
      \sum_{s = 1}^6 V^{(0)\,q\, RL,LR}_{s\,fi} 
  B_0 \left( {m_{\tilde g} ,m_{\tilde q_s } } \right),
\label{selbstenergie}
\end{equation}
satisfying $\Sigma^{q\,RL}_{fi}(0)=\Sigma^{q\,LR *}_{if}(0)$.  In
\eq{selbstenergie} we have diagonalised the squark mass matrix, with the
eigenvalues denoted by $m_{q_s}$. The quantities $V^{q\, RL}_{s\,fi}$
and $V^{q\, LR}_{s\,fi}$ are combinations of the rotation matrices of
quarks and squarks fields and are defined in \eq{abbr} of the appendix,
where also our conventions for the loop functions $B_0$, $C_0$ and $D_0$
are listed.  In the second order of the mass insertion approximation
(neglecting terms with more than one flavour change) \eq{selbstenergie}
becomes (for $f\neq i$)
\begin{eqnarray}
\Sigma^{q\,LR}_{fi} (p^2=0) &=&
\alpha _s (M_{\rm SUSY}) \frac{{2m_{\tilde g} }}{{3\pi }}
  \Big[ 
  \Delta^{\tilde q\, LR}_{fi}
  C_0 \left( m_{\tilde g},  \left[M_{\tilde q}^2\right]_{ff} , 
                           \left[M_{\tilde q}^2\right]_{i+3,i+3}  \right)
  \nn 
&& \;\;
   + \, 
   \Delta^{\tilde q\, LL}_{fi} \, \Delta^{\tilde q\, LR}_{ii} \, 
 D_0 \left( m_{\tilde g},  \left[M_{\tilde q}^2\right]_{ff} , 
                            \left[M_{\tilde q}^2\right]_{ii} , 
                            \left[M_{\tilde q}^2\right]_{i+3,i+3} \right)
  \nn 
&& \;\;
  + \, 
   \Delta^{\tilde q\, LR}_{ff} \, \Delta^{\tilde q\, RR}_{fi} \, 
 D_0 \left( m_{\tilde g},  \left[M_{\tilde q}^2\right]_{ff} , 
                            \left[M_{\tilde q}^2\right]_{f+3,f+3} , 
                            \left[M_{\tilde q}^2\right]_{i+3,i+3} \right)
  \Big].
  \label{mia}
\end{eqnarray}
Even if the flavour-changing elements $\Delta^{\tilde q\, LL}_{fi}$ are
small, the approximation of \eq{mia} breaks down, if $\Delta^{\tilde q\,
  LR}_{jj}$ is of the order of $(M_{jL,R}^{\tilde q})^2$ for either
$j=i$ or $j=f$, i.e.\ for large flavour-diagonal left-right mixing.
This is the default situation for the top squarks and also happens with
the bottom squarks if $\tan\beta$ is large. We therefore work with the
exact formula of \eq{selbstenergie}.

We now collect the results of the diagrams in \fig{fig:W} with the 
simplification that we neglect all small ratios of quark masses such as 
$m_s/m_b$.  One finds
\begin{equation}
  D_{L\, fi} =  \sum\limits_{j = 1}^3 V^{(0)}_{fj} 
                \left[ \Delta U_L^d \right]_{ji},
 \qquad\qquad
  D_{R\, fi} =  \sum\limits_{j = 1}^3 
              \left[ \Delta U_L^{u\dag} \right]_{fj} V^{(0)}_{ji} 
\label{res}
\end{equation}

with the matrices

\begin{equation}
\renewcommand{\arraystretch}{1.4}
\Delta U_L^q \,=\, 
\left( {\begin{array}{*{6}c}
0 & 
{\frac{1}{{m_{q_2 }}} {\Sigma _{12}^{q\,LR} }  } & 
{\frac{1}{{m_{q_3 }}} {\Sigma _{13}^{q\,LR} } }  \\
           {\frac{{ - 1}}{{m_{q_2 }}} {\Sigma _{21}^{q\,RL} }  } & 
0 & 
{\frac{1}{{m_{q_3 }}} {\Sigma _{23}^{q\,LR} } }  \\
           {\frac{{ - 1}}{{m_{q_3 }}} {\Sigma _{31}^{q\,RL} } } & 
{\frac{{ - 1}}{{m_{q_3 }}}{\Sigma _{32}^{q\,RL} }  } & 0
        \end{array}} \right)
\label{DeltaU} 
\end{equation}

In our super-CKM scheme i) the inclusion of the radiative correction is 
equivalent to the use of the tree-level coupling in the $\bar{u}W^+d$ vertex with the replacement

\begin{equation}
        V^{(0)} \to V  =   \left(1+\Delta
          U_L^{u\dag}\right)V^{(0)}\left(1+\Delta U_L^d\right) 
 \label{physv}
\end{equation}

and we identify $V$ with the physical CKM matrix. In the on-shell scheme
ii) the counter-terms $\delta U_L^d=-\Delta U_L^d$ and $\delta
U_L^u=-\Delta U_L^u$ cancel the loops and $V^{(0)}=V$ is maintained. It
is crucial that $1+\delta U_L^q$ is unitary, otherwise the unitarity of
$V$ (and electroweak gauge invariance) would be spoiled
\cite{Denner:1990}. To our one-loop order this means that $\delta U_L^q$
is anti-hermitian.  We can easily verify from \eq{DeltaU} that $\Delta U_L^d$
fulfills this criterion owing to $\Sigma^{q\,RL}_{fi}(0)=[\Sigma^{q\,LR
  }_{if}(0)]^*$. The self-energies do not decouple for $M_{\rm SUSY}\to
\infty$ and, in accordance with the decoupling theorem
\cite{Appelquist:1974tg}, we find that their mere effect is the
renormalisation of the CKM matrix, as implemented in scheme ii).

It is important to stress that the replacement rule in \eq{physv} only
absorbs the effects of the self-energy diagrams of Fig.~\ref{fig:W}
correctly, if both quark lines are external lines. If some $\ov u_f
W^+d_i$ vertex appears in a loop diagram, one or both self-energies are
probed off-shell and one must work with $V^{(0)}$ and must calculate the
loop diagram with the nested self-energy explictly.

We can now understand how to treat self-energies with a top quark in
\eq{DeltaU}: If the top quark appears on the internal line of the right
diagram in \fig{fig:W}, that is $j=3$, the self-energy involved must be
evaluated at $p^2=0$, because the external quark is up or charm. The
unitarity of $V$ now forces us to evaluate $\Sigma^{u\,RL}_{31}$ and
$\Sigma^{u\,RL}_{32}$ at $p^2=0$ as well. Interestingly, from today's
precision data in K and B physics one can determine $V$ from tree-level
data only \cite{Nierste:2005px}. Of course, none of these measurements
involves top decays, so that the values of $V_{ts}$ and $V_{td}$
inferred from these measurements (through unitarity of $V$) indeed
correspond to the definition in \eq{physv}, with self-energies
$\Sigma^{u\,RL}_{3i}$ evaluated at $p^2=0$. While FCNC processes of K
and B mesons involve $V_{ts}$ or $V_{td}$, we cannot determine these CKM
elements from FCNC processes in a model-independent way, because new
particles (in our case squarks and gluinos) will affect the FCNC loops
directly.  Clearly nothing can be learned from measuring the $\ov t W^+
d_i$ couplings (in, for instance, single top production or top decays)
if $M_{\rm SUSY}\gg m_{u_3}=m_t$. However, if $ m_t\sim M_{\rm SUSY}$
any on-shell $t\to s$ or $t\to d$ transition involves

\begin{equation}
  \Delta \sigma^t_i \equiv   
  \frac{\Sigma^{u\,RL}_{3i}(m_t^2)-\Sigma^{u\,RL}_{3i}(0)}{m_t} 
  \qquad\qquad
  \mbox{with $i=1$ or 2}. \label{topself}
\end{equation}

Here the first self-energy enters the calculated $\ov t W^+ d_i$ process
explicitly, while $\Sigma^{u\,RL}_{3i}(0)$ stems from the relationship
between $V$ and $V^{(0)}$. $\Delta \sigma^t_i$ decouples as
$m_t^2/M_{\rm SUSY}^2$, but can be sizable for ${\cal O}(200\, \gev)$
superpartners, since it involves poorly-constrained FCNC squark mass
terms.  We conclude that the flavour structure of tree-level top
couplings can help to study new physics entering chirality-flipping
self-energies, while this effect is unobservable in charged-current
processes of light quarks: Here the chirality-flipping self-energies
merely renormalise the CKM matrix; the physical effect in a
charged-current processes with external quark $q$ is suppressed by a
factor of $m_q^2/M_{\rm SUSY}^2$. The experimental signature would be an
apparent violation of CKM unitarity, since the measured value of
$V_{ts}$ or $V_{td}$ would be in disagreement with the value inferred
from CKM unitarity. Unitarity is restored, once the correction $\Delta
\sigma^t_i$ is taken into account.

Since inverse quark masses enter \eq{DeltaU}, we must address the
proper definition of these masses in the presence of ordinary QCD
corrections. If we worked in the decoupling limit and calculated the
diagrams of \fig{fig:dec}, we would encounter the $\ov{\rm
  MS}$-renormalised Yukawa couplings evaluated at the renormalisation
scale $Q=M_{\rm SUSY}$, at which the heavy SUSY particles are integrated
out. Translating that result into the language of Sect.~\ref{sect:2}
amounts to the evaluation of the inverse quark masses in the 
$\ov{\rm MS}$ scheme at $Q=M_{\rm SUSY}$. One can derive this (somewhat
surprising) result entirely in the diagrammatic language of
Sect.~\ref{sect:2}, by studying QCD corrections to the diagrams of
\fig{fig:W} \cite{hns}. The first element in this proof is the
observation that e.g.\ $\Sigma _{fi}^{q\,LR}$, viewed as the Wilson
coefficient of the two-quark operator $\ov q_f P_R q_i$, renormalises in
the same way as the quark mass, so that the ratios  
$\Sigma _{fi}^{q\,LR}/m_{q_i}$ in \eq{DeltaU} are independent of $Q$. 
Since the SUSY parameters entering $\Sigma _{fi}$ are defined at the
high scale $Q=M_{\rm SUSY}$, our constraints derived in the next section
will involve $m_{q_i}(M_{\rm SUSY})$. The second element in the proof 
is the explicit analysis of gluonic corrections to the diagrams of
\fig{fig:W}. While at intermediate steps a quark pole mass enters
through the Dirac equation $\ps q_i = m_{q_i}^{\rm pole} q_i$, gluonic
self-energies add to $m_{q_i}^{\rm pole}$ in such a way that the final
result only involves the properly defined $\ov{\rm MS}$ mass $m_{q_i}$
\cite{hns}. 
   
We close this section by recalling the relationship between the Yukawa
matrices ${\bf Y}^q = \mbox{diag}\, ( Y^{q_1}, Y^{q_2}, Y^{q_3})$ and the
quark masses \cite{Hall:1993gn,Carena:1999}:

\begin{equation}
Y^{q_i} \quad=\quad \frac{m_{q_i}}{v_q \, (1 + \Delta_{q_i}) } \quad=\quad  \frac{m_{q_i}-\Sigma^{q\,LR}_{ii,\,A}}{v_q \, (1 +\frac{\Sigma^{q\,LR}_{ii,\,\mu}}{m_{q_i}} ) }.\label{mqdel}
\end{equation}

In \eq{mqdel} we have used the fact that $\Sigma^{q\,LR}_{ii}$ can be
decomposed into $\Sigma^{q\,LR}_{ii,\,A}+\Sigma^{q\,LR}_{ii,\,\mu}$ if the
physical squark masses are chosen as input parameters.
$\Sigma^{q\,LR}_{ii,\,\mu}$ is proportional to $\mu\,Y^{q_i}$ and
$\Sigma^{q\,LR}_{ii,\,A}$ is proportional to $A^{q}_{ii}$. If we neglect the
A-terms \eq{mqdel} reduces to the expression of \cite{Carena:1999} for
down-type quarks. For a detailed discussion of the relation between the Yukawa
matrices and the quark masses with different choices of input parameters see
\cite{hns}.

\eq{mqdel} holds in the super-CKM scheme i), which has the advantage that no
FCNC Yukawa couplings occur.  In the on-shell scheme ii) the rotations of the
quark fields in flavour space lead to the loop-induced finite FCNC Yukawa
couplings of
Refs.~\cite{Hamzaoui:1998nu,Babu:1999hn,Isidori:2002qe,Buras:2002vd}. In the
super-CKM scheme these effects are reproduced from diagrams with
flavour-diagonal Yukawa couplings and FCNC self-energies. Finally note that
$\Delta_{q_i}$ can be complex, so that the entries of ${\bf Y}^{q}$ (and
$m_q^{(D)}={\bf Y}^{q} v_q$ entering the squark mass matrices in \eq{upsckm})
are not necessarily real.

\section{Numerical Analysis}\label{sect:3}

Large accidental cancellations between the SM and supersymmetric
contributions are, as already mentioned in the introduction, unlikely
and from the theoretical point of view undesirable. Requiring the
absence of such cancellations is a commonly used fine-tuning argument,
which is also employed in standard FCNC analyses of the $\delta^{q \,XY}
_{ij}$'s \cite{Hagelin:1992, Gabbiani:1996,Ciuchini:1998ix,
  Borzumati:1999, Becirevic:2001,Silvestrini:2007,Ciuchini:2007cw}.
Analogously, we assume that the corrections due to flavour-changing SQCD
self-energies do not exceed the experimentally measured values for the
CKM matrix elements quoted in the Particle Data Table (PDT)
\cite{Yao:2006px}. To this end we set the tree-level CKM matrix
$V^{(0)}$ (working in our super-CKM scheme i)) equal to the unit matrix
and generate the measured values radiatively.  For simplicity, we
consider the up and down sector separately.  We parameterise the squark
mass matrix according to \eq{massmatrix}.  We choose all diagonal
elements of the squark mass matrix to be equal and denote them by
$m_{\tilde{q}}^2$. For small off-diagonal elements $m_{\tilde{q}}$ is in
good approximation equal to the physical squark mass.

The quantities $\delta^{q\, LR} _{ij}$ and $\delta^{q\, RL} _{ij}$ as
defined in \eq{defde} violate SU(2) and are therefore proportional to
the vacuum expectation value (vev) of a Higgs field. This means that if
one scales the SUSY spectrum by a common factor $a$, the
chirality-flipping elements $\delta^{q\, LR,RL} _{ij}$ will scale like
$1/a$ rather than staying constant as one might naively expect for a
dimensionless quantity.  This leads to somewhat counter-intuitive bounds
on $\delta^{q\,LR} _{ij}$ and $\delta^{q\,RL} _{ij}$, which become
stronger for a larger scale parameter $a$. However, the inferred bounds
on the trilinear terms $A^{q}_{ij}$ stay constant for $a \gg 1$ as
expected for a bound derived from a non-decoupling quantity.  Since it
is customary in nearly all treatments of non-minimal flavour violation
to constrain $\delta^{q\,XY} _{ij}$ \cite{Hagelin:1992, Gabbiani:1996,
  Ciuchini:1998ix,
  Borzumati:1999,Becirevic:2001,Silvestrini:2007,Ciuchini:2007cw}, we
will also follow this convention rather than quoting constraints on
$A^{q}_{ij}$.

We use the standard parameterisation for the CKM matrix and quark
masses defined in the $\overline{\rm MS}$ scheme with the central PDT
values \cite{Yao:2006px}. As discussed at the end of Sect.~\ref{sect:2},
the masses enter the loop contributions to $V$ in \eq{DeltaU} at the
renormalisation scale $Q=M_{\rm SUSY}$, at which the self-energies are
calculated.   

Because of the V--A structure of the W vertex only self-energies
$\Sigma^{q\,LR}_{fi}$ with $f<i$ enter the renormalisation of the CKM
matrix in the approximation $m_{q_f}\,=\,0$ (see \eq{DeltaU}). This
implies that for $f<i$ only $\delta^{q\,LR}_{fi}$ is constrained in the
leading order of the mass insertion approximation in \eq{mia}. In order
to constrain $\delta^{q\,LL}_{fi}$ or $\delta^{q\,RR}_{fi}$ an
additional insertion of some $\delta^{q\,LR}_{jk}$ is needed.  Even
three insertions are needed to involve $\delta^{q\,RL }_{fi}$. These
diagrams with multiple insertions of $\delta^{q\,XY }_{jk}$ can give
useful bounds on one of these quantities only if the $\delta^{q\,LR
}_{jk}$ providing the needed chirality-flip is large. The only candidate
is the flavour-diagonal mass insertion $\delta^{q\,LR,RL }_{jj}$ and the
corresponding contributions to $\Sigma^{q\,LR}_{fi}$ can be read off
from the last two terms in \eq{mia}.  Indeed, we find useful bounds on
$|\delta^{d\, LL}_{13}|$ in the large-$\tan\beta$ scenario, where
$\delta^{d\,LR }_{33}/m_{d_3}$ is large.  The analogous contributions
involving $\delta^{d\,RR }_{ij}$ are suppressed with respect to those
with $\delta^{d\, LL}_{ij}\delta^{d\,LR }_{jj}$ by a small ratio of
quark masses. The upper limits on $\delta^{q\,RR }_{fi}$ and
$\delta^{q\,RL }_{fi}$ from vacuum stability \cite{Casas:1995}, electric
dipole moments and FCNC processes are stronger than ours.

We use the following input parameters \cite{Yao:2006px}:
\begin{eqnarray}
\ov{m}_s (2\,\gev) &=& 0.095\, \gev, \qquad  \qquad  
\ov{m}_c (\ov{m}_c) \;=\; 1.25\, \gev, \nn
\ov{m}_b (\ov{m}_b) &=& 4.2\, \gev, \qquad \qquad
~~~~\ov{m}_t (\ov{m}_t) \;=\; 166 \, \gev, \nn
|V_{us}| &=& 0.227,\, \qquad \quad  
|V_{ub}| \;=\; 0.00396,\, \qquad\quad  
|V_{cb}| \;=\; 0.0422.
\end{eqnarray}

\subsection{Down-sector}
We present our bounds on $|\delta^{d\, LR}_{ij}|$ and $|\delta^{d\, LL}_{ij}|$
in Sects.~\ref{sect:dlr} and \ref{sect:dll}, respectively.

\subsubsection{Constraints on 
               $\mathbf{|\delta^{d\, LR}_{ij}|}$}\label{sect:dlr}

\paragraph{Constraints from $\mathbf{V_{us}}$,$\mathbf{V_{cd}}$:}

$V_{us}$ and $V_{cd}$ are experimentally well known. Their absolute
values are nearly equal and they have opposite sign in the standard
parameterisation, which is respected by the corrections \eq{DeltaU}.
Fig.~\ref{CKMd12} shows the dependence of the constraints on the squark
mass with different ratios $m_{\tilde{g}}/m_{\tilde{q}}$. In the
approximation $m_{d_{1}}=0$, only $\delta^{d\, LR}_{12}$ is constrained.
Looking at the dependence on the gluino mass (Fig.~\ref{CKMd12gluino}),
it is interesting to note that there is a minimum for
$m_{\tilde{g}}\approx 1.5m_{\tilde{q}}$.

\begin{nfigure}{tbp}
\includegraphics[width=1\textwidth]{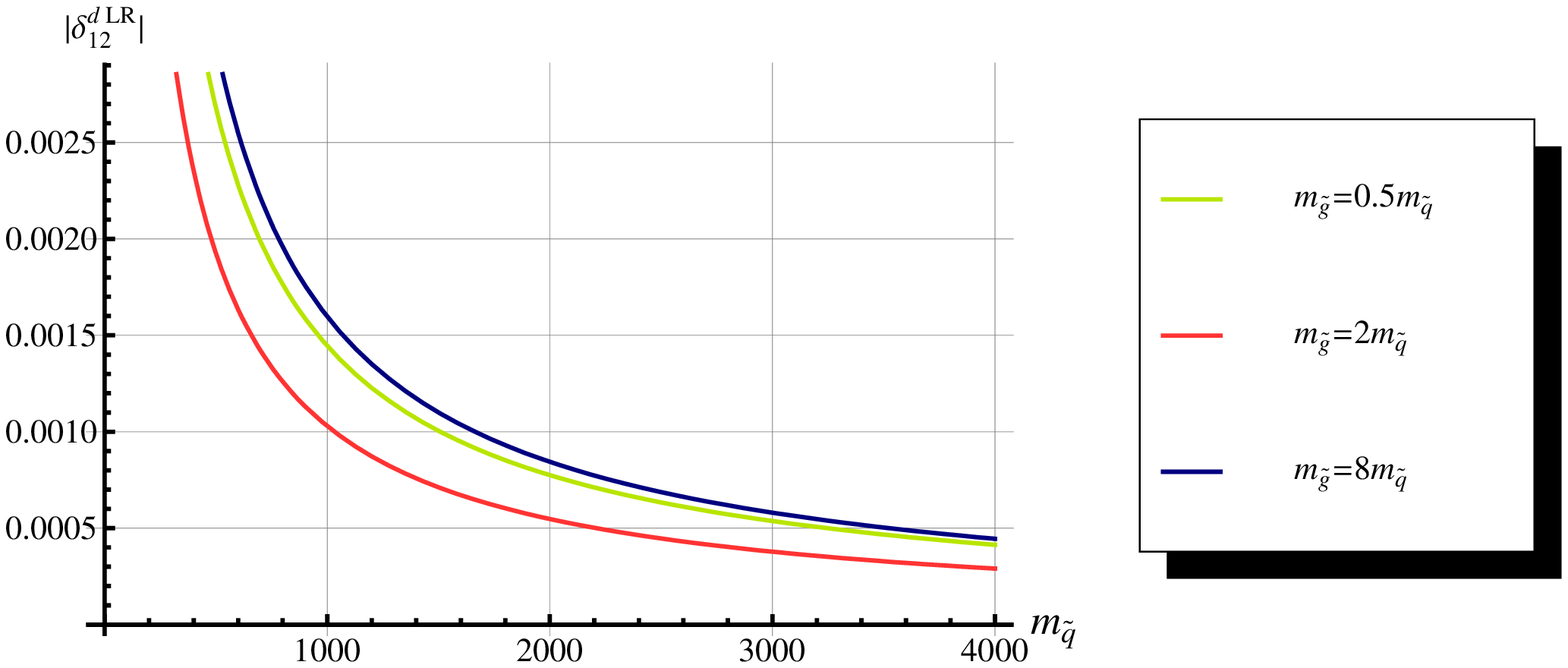}
\caption{Constraints on $|\delta^{d\, LR}_{12}|$ from $V_{us}$ or
  $V_{cd}$ as a function of the squark mass. The constraints become
  stronger with growing $M_{\rm SUSY}$, because $\ds \delta^{q\,
    LR}_{ij}\propto \frac{v}{M_{\rm SUSY}}$. \label{CKMd12}}
\end{nfigure}

\begin{nfigure}{tbp}
\includegraphics[width=1\textwidth]{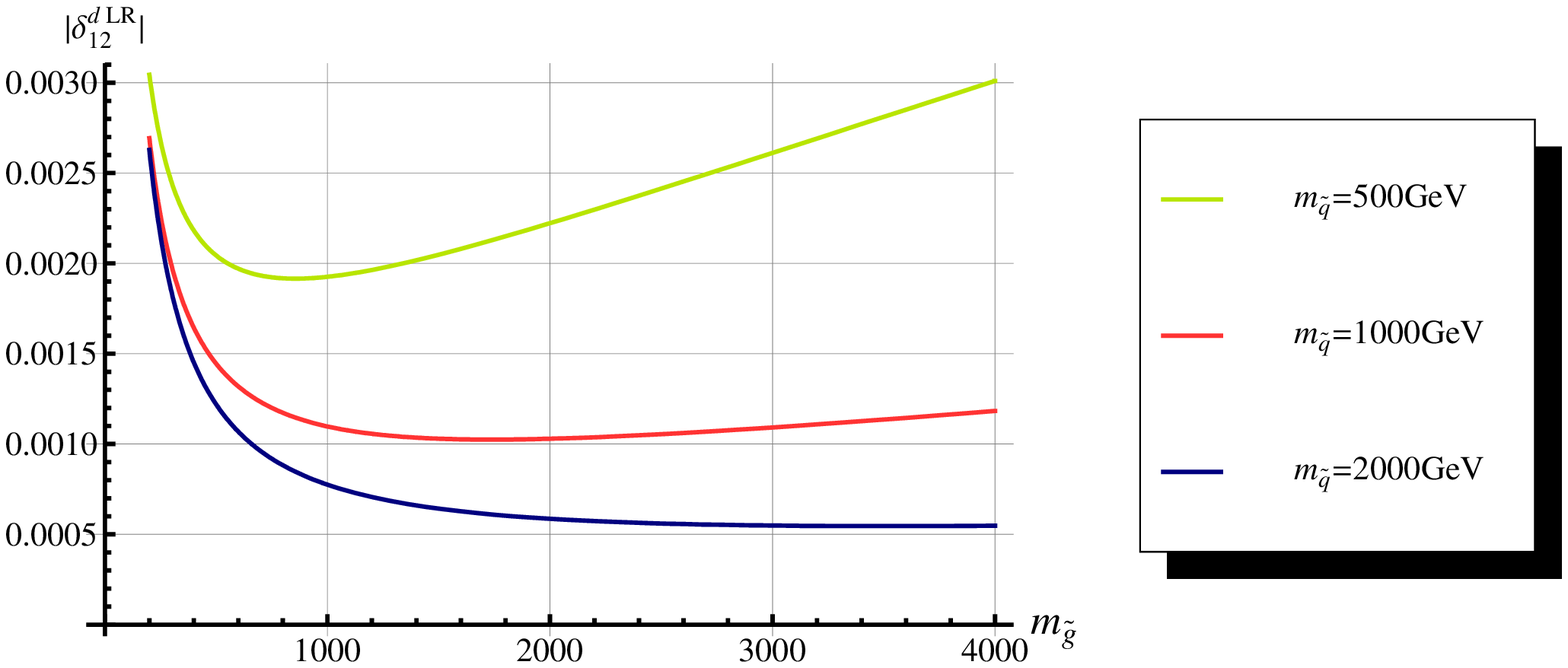}
\caption{Constraints on $\delta^{d\, LR}_{12}$ from $V_{us}$ (or $V_{cd}$) as
  a function of the gluino mass.\label{CKMd12gluino}}
\end{nfigure}

\paragraph{Constraints from $\mathbf{V_{cb}}$,$\mathbf{V_{ts}}$:}
In this case, the situation is nearly the same as in the case of $V_{us}$,
except that the constraints are weaker (see Fig.~\ref{CKMd23}), because 
$m_b$ is much larger than $m_s$.  $|V_{ts}|$ is essentially fixed by the
measured value of $|V_{cb}|$ through CKM unitarity.

\begin{nfigure}{tbp}
\includegraphics[width=1\textwidth]{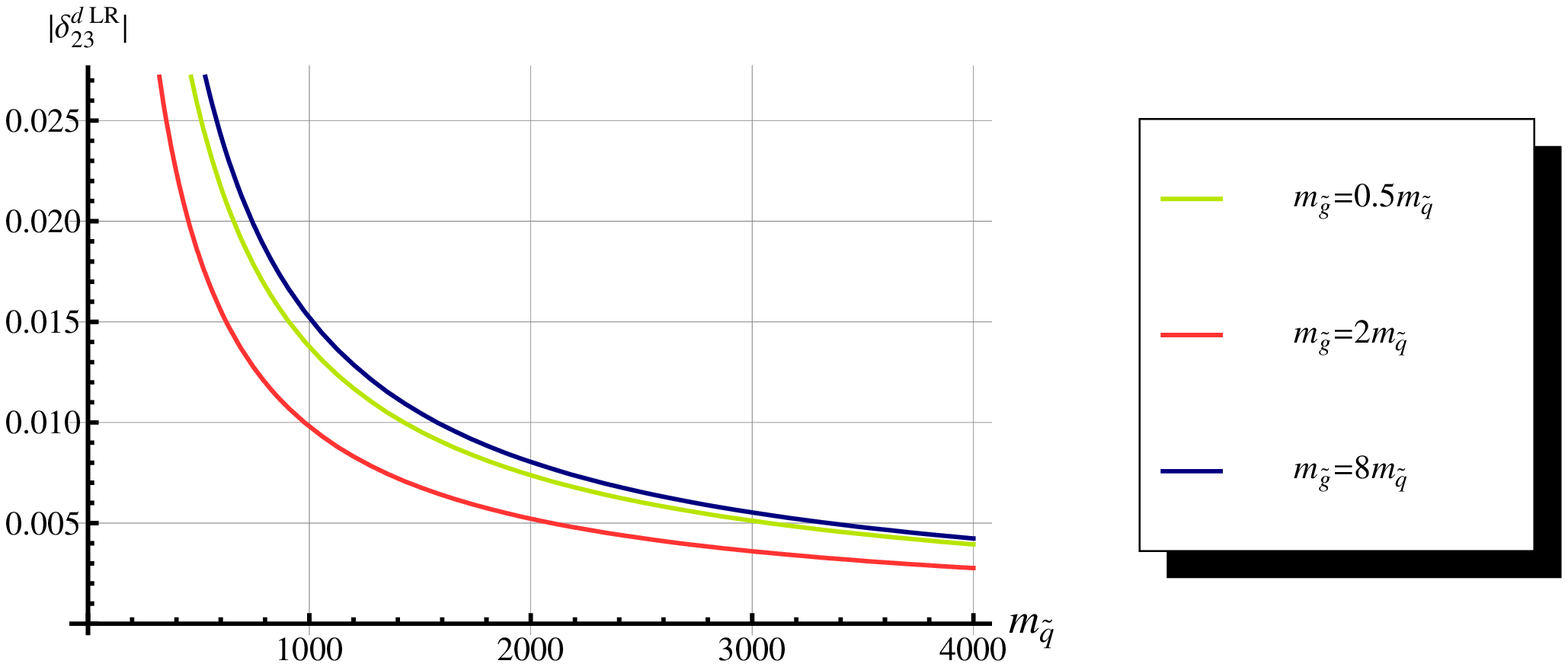}
\caption{Constraints on $\delta^{d\, LR}_{23}$ from $|V_{cb}|$ (or
  $|V_{ts}|$) as a function of the squark mass.\label{CKMd23}}
\end{nfigure}

\begin{nfigure}{tbp}
\includegraphics[width=1\textwidth]{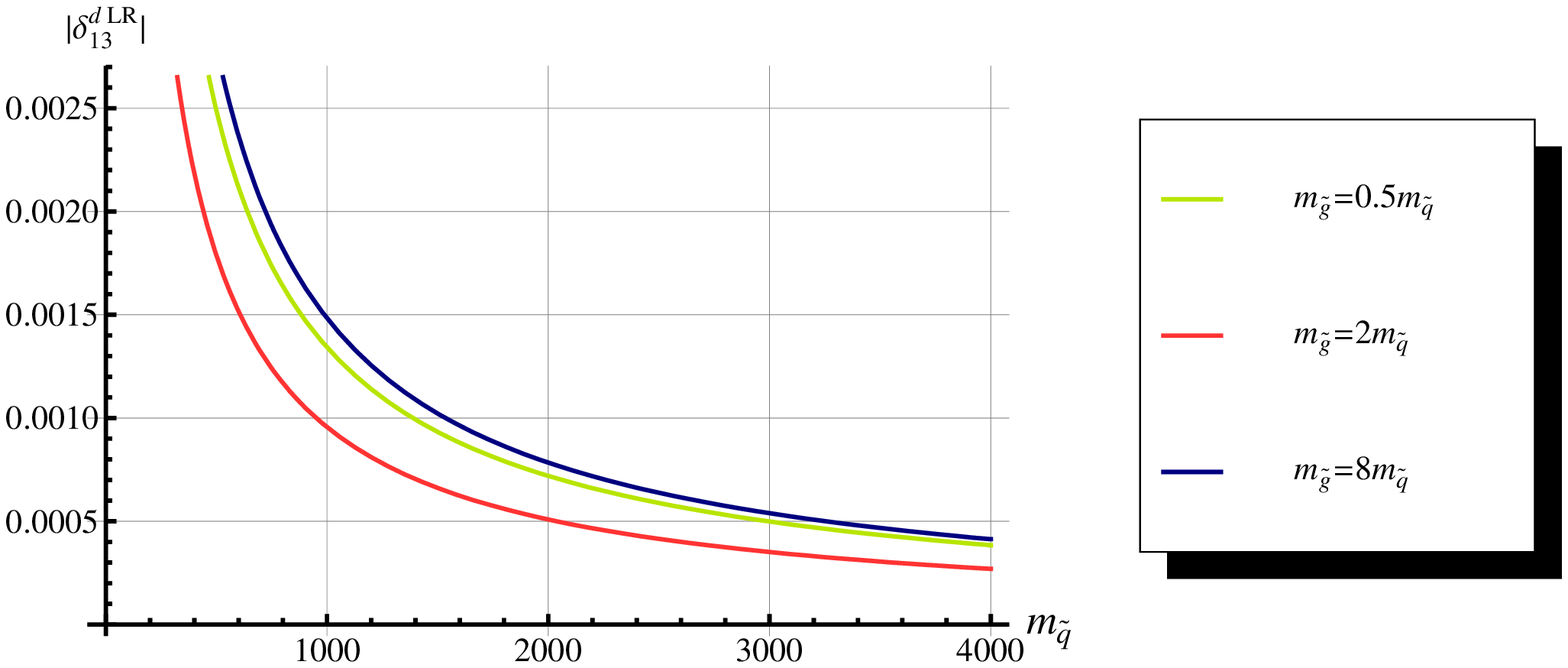}
\caption{Constraints on $\delta^{d\, LR}_{13}$ from 
$|V_{ub}|$ as a function of the squark mass.\label{CKMd13}}
\end{nfigure}

\begin{nfigure}{tbp}
\includegraphics[width=1\textwidth]{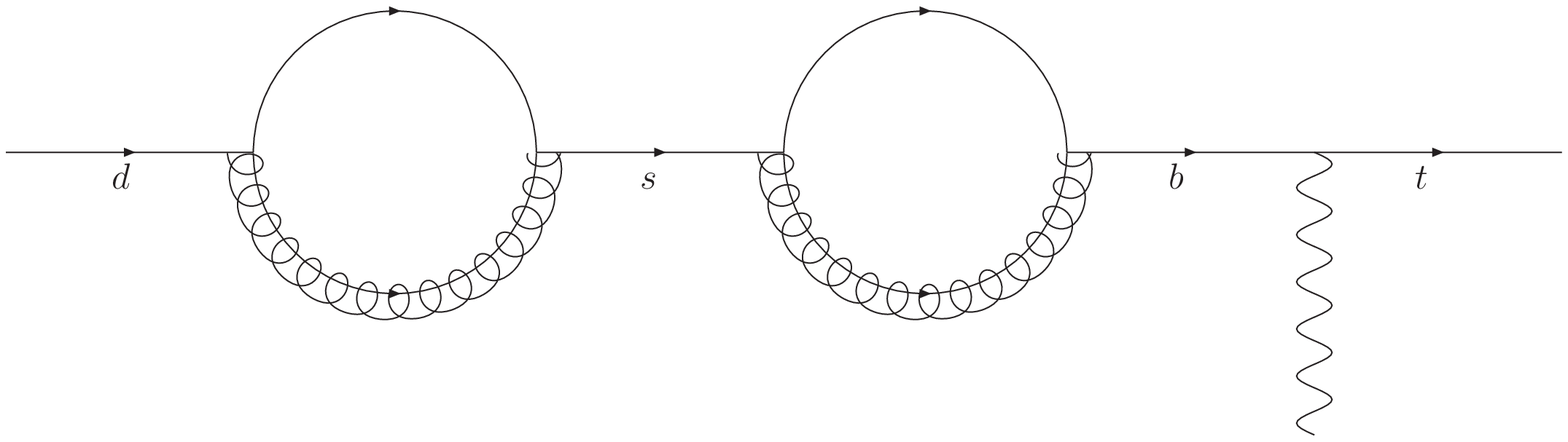}
\caption{Two-loop correction to $V_{td}$.\label{fig:dsbW}}
\end{nfigure}

\begin{nfigure}{tbp}
\includegraphics[width=1\textwidth]{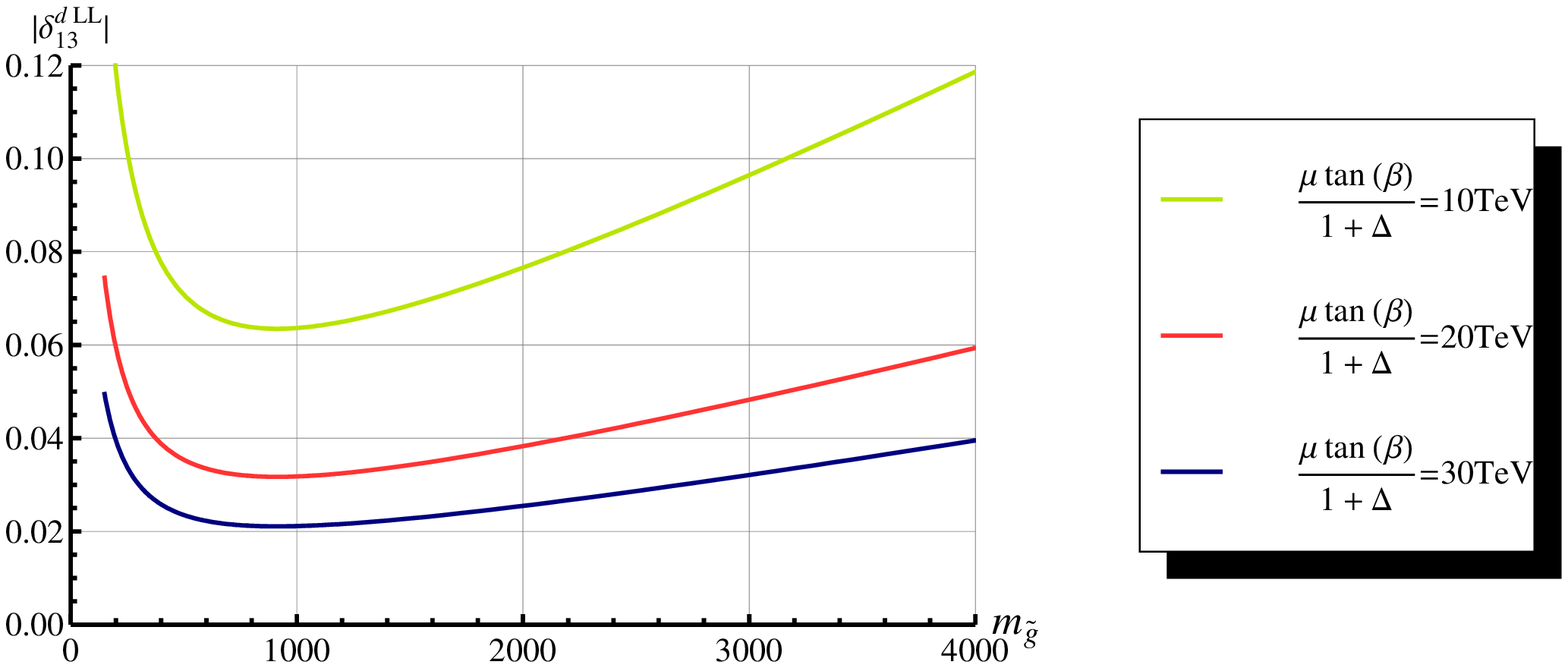}
\caption{Constraints on $\delta^{d\, LL}_{13}$ from $|V_{ub}|$  as a
  function of the gluino mass for different values of 
  $\mu\, \tan\beta/(1+\Delta_b)$.\label{CKMd13LL}}
\end{nfigure}

\paragraph{Constraints from $\mathbf{V_{ub}}$:}
The last pair of of CKM elements to be discussed is $(V_{ub},V_{td})$.
In this case $|V_{ub}|$ does not fix $|V_{td}|$, because $|V_{td}|$ is
largely affected by the CKM phase.  Now $|V_{ub}|$ is experimentally
better known than $|V_{td}|$, because $V_{td}$ is extracted from FCNC
loop processes by comparing the corresponding experimental result with
the SM prediction. In beyond-SM scenarios, this is not a valid
procedure anymore, because the new particles will alter the FCNC loop
processes.  Therefore we can only exploit the constraint from
$|V_{ub}|$.  The anti-hermiticity of $\Delta U_L^d$ in \eq{DeltaU}
implies that $\delta^{d\, LR}_{13}$ gives a negative contribution of the
same size to $V_{td}$.  This cannot be the whole contribution, since
$|V_{ub}|\neq |V_{td}|$ and in the standard CKM parameterisation $\real
V_{td}$ and $\real V_{ub}$ are both positive.  The hierarchical
structure of the CKM matrix is responsible for a second contribution:
Since $V_{ub}$ and $V_{td}$ are of third order in the Wolfenstein
parameter $\lambda$, the two-loop process in Fig.~\ref{fig:dsbW}
involving the loop-contributions to $V_{us}\propto \lambda$ and
$V_{cb}\propto \lambda^2$ is important as well.  This diagram adds a
contribution of $V_{us}V_{cb}=0.0088$ to $V_{td}$. Together with the
one-loop contribution from $\delta^{d\, LR}_{13}$, this yields the
correct value for $V_{td}$. We stress that this does not imply any
additional constraint on the SUSY parameters entering the self-energies,
to order $\lambda^3$ we just reproduced a unitarity relation of the CKM
elements: $V_{td}=-V_{ub}^* + V_{cd} V_{ts} \simeq - V_{ub}^* - V_{us}^*
V_{ts}$, which (with insertion of $V_{tb}\simeq V_{ud}\simeq 1$) equals
the product of the first and third rows of the CKM matrix.

The last possible constraint is the phase of the CKM matrix, which one
could infer from $\gamma=\arg(-V_{ub}^*V_{ud}/(V_{cb}^* V_{cd})
)$. However, since $\gamma$ is large, no fine-tuning argument can be
applied to derive bounds. Only in a given scenario of radiatively
generated CKM elements, the measured value of $\gamma$ can be used to 
derive a constraint on the complex phases in the mass matrix of 
\eq{massmatrix}.

\subsubsection{Constraints on $\mathbf{\delta^{d\, LL}_{ij}}$}\label{sect:dll}

In the presence of large chirality-flipping flavour-diagonal elements in the
squark mass matrix, also $\delta^{q\, LL}_{ij}$ can be constrained.  This is
the case for large $A^q_{jj}$ terms or (if $q=d$) for a large value of
$\mu\,\tan \beta $. Here we only consider the second possibility, which is
widely studied in the literature. The strongest constraints are obtained for
$\delta^{d\, LL}_{13}$, because $V_{ub}$ is the smallest entry of the CKM
matrix.  We have included the correction term $\Delta_b$ of \eq{mqdel} in
our analysis.  Our result is shown in Fig.~\ref{CKMd13LL}. Our constraint is
compatible with the experimental bound on $Br(B_d \to \mu^+ \mu^-)$
for values of $\tan\beta$ around 30 or below \cite{Kane:2003wg}.

We next discuss the constraint on $\delta^{d\, LL}_{23}$: It is clear that our
bound will be looser by a factor of $|V_{cb}/V_{ub}|$. Furthermore, for large
$\tan \beta$ and typical values of the massive SUSY parameters we find $Br(B_s
\to \mu^+ \mu^-)$ more constraining. To find bounds on $\delta^{d\, LL}_{23}$
from $|V_{cb}|$ which comply with $Br(B_s \to \mu^+ \mu^-)$ we need a smaller
value of $\tan\beta$ around 20 and therefore a quite large value for $\mu$, 
if the masses of the non-standard Higgs bosons are around 500$\,\gev$. We
do not include a bound on $\delta^{d\, LL}_{23}$ in our table of results in
Sect.~\ref{sec:comp}.
  
\begin{nfigure}{tbp}
\includegraphics[width=1\textwidth]{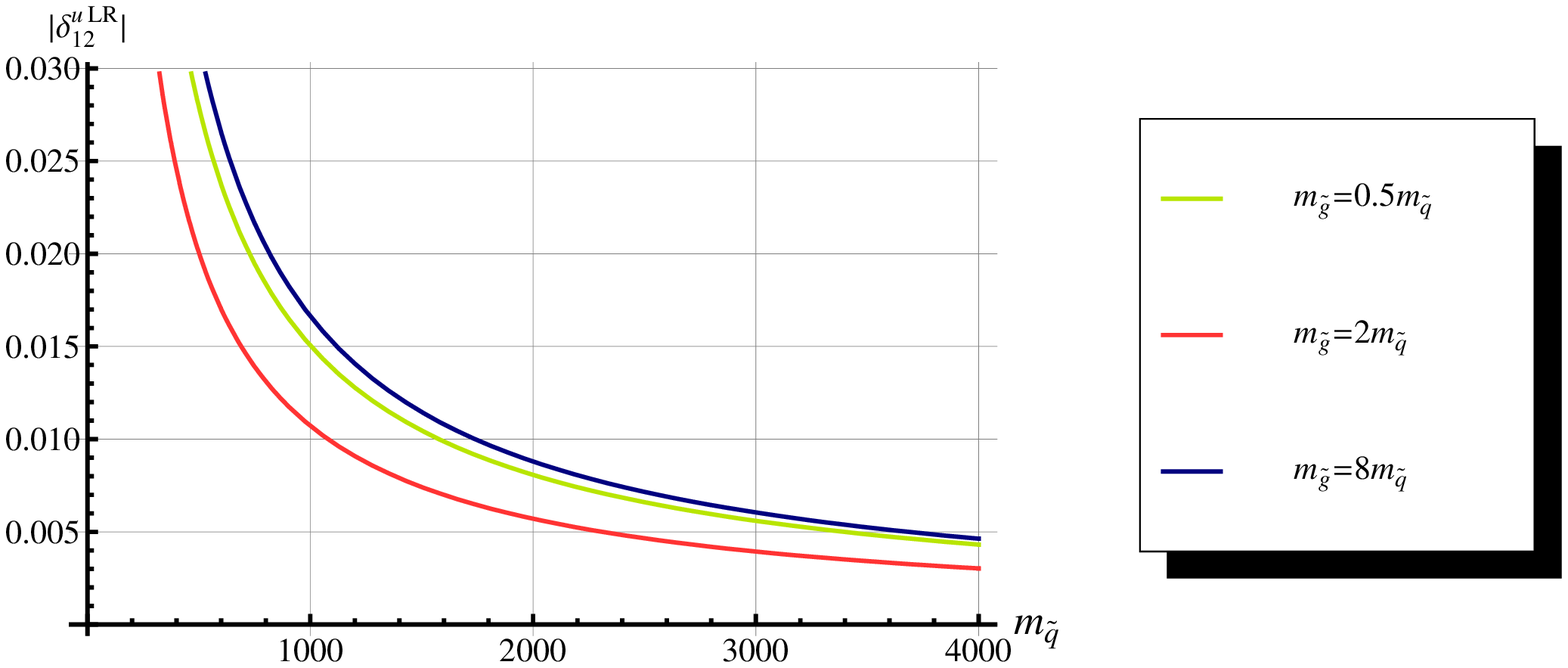}
\includegraphics[width=1\textwidth]{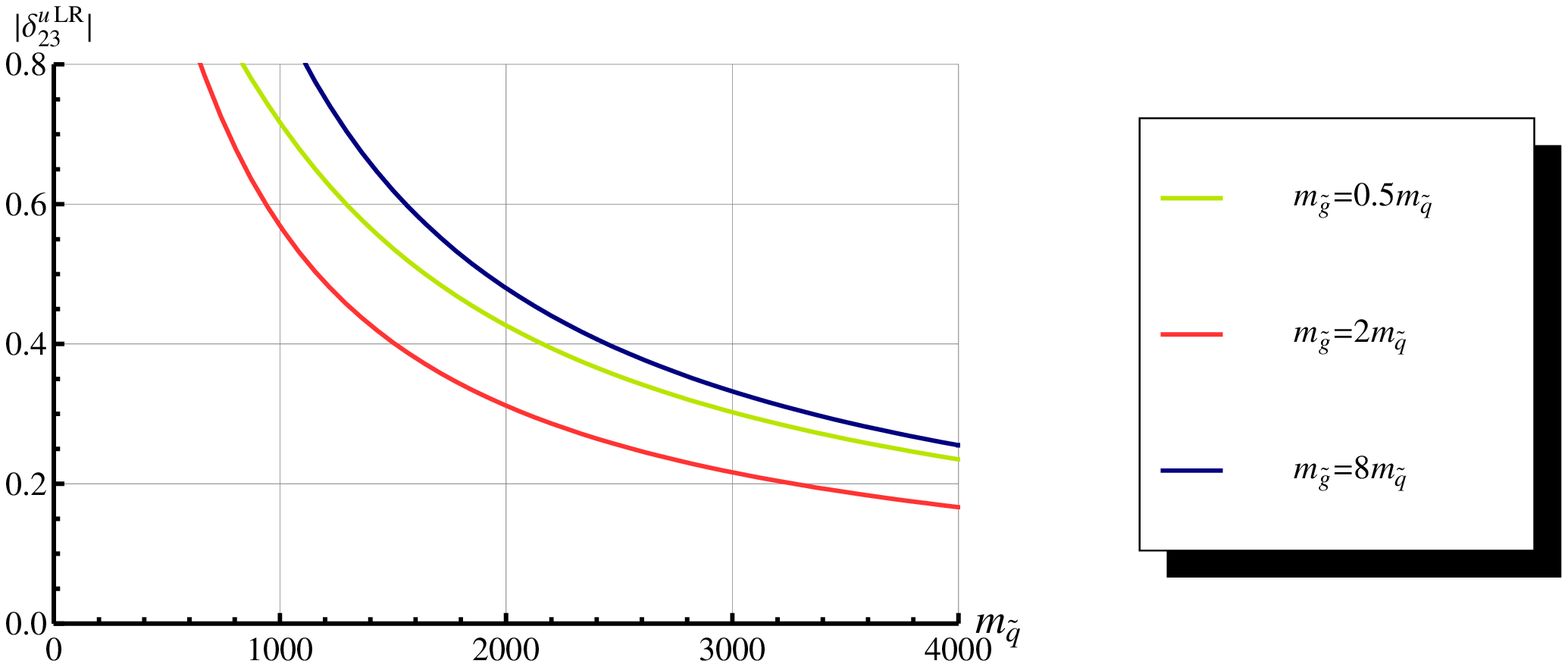}
\includegraphics[width=1\textwidth]{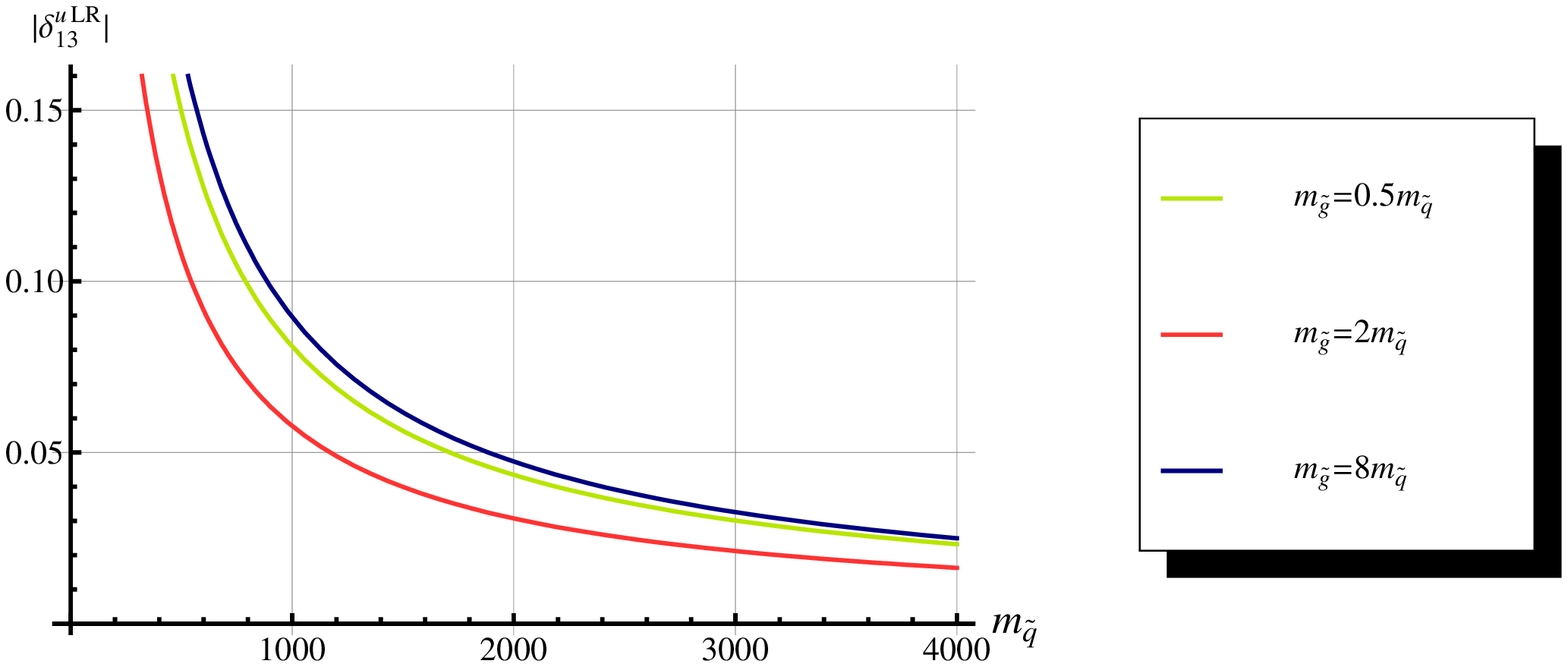}
\caption{Constraints on $\delta^{u\, LR}_{12}$, $\delta^{u\, LR}_{23}$
  and $\delta^{u\, LR}_{13}$ as a function of the squark mass for
  different ratios 
  of $m_{\tilde g}/m_{\tilde q}$.\label{CKMu13}\label{usector}}
\end{nfigure}

\subsection{Up-sector}

In the up-sector, everything is in straight analogy to the down-sector.  The
only difference is that the constraints are weaker because of the
larger charm and top masses. But the upper bounds are still restrictive,
except the ones obtained from $V_{ts}$ or $V_{cb}$ (see Fig.~\ref{usector}).
Remarkably, we now have a powerful constraint on $\delta^{u\, LR}_{13}$ from
the second diagram in Fig.~\ref{fig:W} with $q_{u_f}=u$ and $q_{d_i}=b$.

\subsection{Comparison with previous bounds}\label{sec:comp}
In this section we compare our bounds with those in the literature, derived
from FCNC processes \cite{Ciuchini:1998ix, Becirevic:2001, Ciuchini:2007cw,
  Silvestrini:2007} and vacuum stability (VS) bounds \cite{Casas:1995}. We
take $M_{\rm SUSY}=\sqrt{\left[M_{\tilde
      q}^2\right]_{ss}}=m_{\tilde{g}}=1000\, \gev$:

\begin{center}
\renewcommand{\arraystretch}{1.4}
\begin{tabular}{cllll}
quantity   & our bound   & \multicolumn{2}{l}{bound from FCNC's}    
& bound from VS \cite{Casas:1995}\\\hline
$|\delta^{d\,LR}_{12} |$ & $\leq 0.0011$   & $\leq 0.006$ & 
$K$ mixing \cite{Ciuchini:1998ix}   & $\leq 1.5\, \times \,10^{-4}$  \\
$|\delta^{d\,LR}_{13} |$ & $\leq 0.0010$   & $\leq 0.15$ & 
$B_d$ mixing \cite{Becirevic:2001}   & $\leq 0.05$  \\
$|\delta^{d\,LR}_{23} |$ & $\leq 0.010$        & $\leq 0.06$ & 
 $B\rightarrow X_s\gamma; X_s l^+l^-$ \cite{Silvestrini:2007}   & 
$\leq 0.05$  \\
$|\delta^{d\,LL}_{13} |$ & $\leq 0.032$ & $\leq 0.5$ & 
$B_d$ mixing \cite{Becirevic:2001}    & $-$  \\
$|\delta^{u\,LR}_{12} |$ & $\leq 0.011$  & $\leq 0.016$  & 
$D$ mixing \cite{Ciuchini:2007cw}  & $\leq 1.2\, \times \,10^{-3}$  \\
$|\delta^{u\,LR}_{13} |$ & $\leq 0.062$       & \multicolumn{2}{l}{~~~~~~~~---}
& $\leq 0.22$  \\
$|\delta^{u\,LR}_{23} |$ & $\leq 0.59$        & \multicolumn{2}{l}{~~~~~~~~---}
& $\leq 0.22$  \\                      
\end{tabular}
\end{center}
Our value for $\delta^{d\,LL}_{13}$ is calculated with $\ds \frac{\mu\,
  \tan\beta}{1+\Delta_b}\,=\, 20 \,\tev$.  The quoted bound on
$\delta^{d\,LR}_{23}$ from $b\to s \gamma$ and $B\to X_s \,l^+l^-$ has been
rescaled by an approximate factor of 3 from the value quoted for $M_{\rm
  SUSY}\,=\,350 \,\gev$ in Ref.~\cite{Silvestrini:2007}. The VS bounds on
$\delta^{u\,LR}_{ij}$ have also been obtained by scaling the quoted values for
$M_{\rm SUSY}\,=\,500 \,\gev$ of Ref.~\cite{Casas:1995} by a factor of $1/2$.
The VS bounds on $\delta^{u\,LR}_{13}$ and $\delta^{u\,LR}_{23}$ are obtained
by multiplying the bound on $\delta^{u\,LR}_{12}$ with $m_t/m_c$.  FCNC
effects are decoupling and scale as $1/M_{\rm SUSY}^2$, but the constraints on
$\delta^{q\,LR}_{ij}$ are proportional to $M_{\rm SUSY}$ rather than $M_{\rm
  {SUSY}}^2$, because the definition of $\delta^{q\,LR}_{ij}$ involves a
factor of $v/M_{\rm SUSY}$.  Both our constraints and the VS bounds on the
trilinear SUSY-breaking terms are independent of $M_{\rm SUSY}$ (i.e.\ 
non-decoupling), so that the bounds on $\delta^{q\,LR}_{ij}$ scale like
$1/M_{\rm SUSY}$. We conclude that all our bounds on $\delta^{d\,LR}_{ij}$ are more restrictive
than those  from FCNC processes for $M_{\rm SUSY}\,\geq \, 500
\,\gev$, and our bound on $\delta^{u\,LR}_{12}$ is stronger than 
the quoted FCNC bound for $M_{\rm SUSY}\,\geq \, 900 \,\gev$. 

Substantially stronger bounds than ours are only listed for the VS bounds on
$|\delta^{u\,LR}_{23} |$, $|\delta^{u\,LR}_{12}|$ and $|\delta^{d\,LR}_{12}|$.
However, the VS bounds related to the latter two quantities are of the form
\bea%
A^q_{12} & < & Y^{q_2} \, f, 
\eea%
where $f={\cal O}(M_{\rm SUSY})$ depends on other massive parameters of the
scalar potential.  The bounds are obtained by studying the scalar potential at
tree level and $Y^{q_2}$ enters the analysis through the quartic coupling of
strange squarks to Higgs bosons. The smallness of $Y^{q_2}$ makes this
coupling sensitive to large loop corrections and the quoted bounds have to be
considered as rough estimates at best.  Our results for $\delta^{q\,LR}_{12}$
rest on a firmer footing.

\subsection{Supersymmetry breaking as the origin of flavour?}\label{sect:sof} 
The smallness of the Yukawa couplings of the first two generations (and
possibly also of the bottom and tau couplings) suggest the idea that Yukawa
couplings are generated through radiative corrections \cite{Weinberg:1972ws}.
In the context of supersymmetric theories these loop-induced couplings arise
from the diagrams of \fig{fig:dec} or, in our approach, of \fig{fig:W}
\cite{Buchmuller:1982ye}.  The B factories have confirmed the CKM mechanism of
flavour violation, leaving little room for new sources of FCNC's.  This
seriously challenges the idea that flavour violation stems from the same
source as supersymmetry breaking.  The surprising discovery of
Sect.~\ref{sect:3} is the finding that this idea is still viable, with SUSY
masses well below 1 TeV, if the sources of flavour violation are the trilinear
terms $A^q_{ij}$. (Note that the VS bound on $|\delta^{u\,LR}_{23} |$ poses
no problem, because one can generate $V_{cb}$ entirely from
$\delta^{d\,LR}_{23}$.)  Of course, the heaviness of the top quark requires a
special treatment of $Y^t$ and the successful bottom-tau Yukawa unification
suggest to keep tree-level Yukawa couplings for the third generation. This
scenario has been studied in Ref.~\cite{Ferrandis:2004}, where possible
patterns of the dynamical breaking of flavour symmetries are discussed.

In the modern language of Refs.~\cite{cg,agis} the global $[U(3)]^3$ flavour
symmetry\footnote{We only discuss the quark sector here.}  of the gauge sector
is broken to $[U(2)]^3 \times U(1)$ by the Yukawa couplings of the third
generation. Here the three $U(2)$ factors correspond to rotations of the
left-handed doublets and the right-handed down--type and up--type singlets of
the first two generations in flavour space, respectively.  That is, our
starting point is ${\bf Y^u}=\mbox{diag}\,(0,0,Y^t) $, ${\bf
  Y^d}=\mbox{diag}\,(0,0,Y^b) $.  We next assume that the soft breaking terms
${\Delta _{ij}^{\tilde{q}\,LL} }$ and ${\Delta _{ij}^{\tilde{q}\,RR} }$
possess the same flavour symmetry as the Yukawa sector, which implies that
${\Delta^{\tilde{q}\,LL} }$ and ${\Delta^{\tilde{q}\,RR} }$ are diagonal
matrices with the first two entries being equal. That is, flavour universality
holds for the first two generations. From a model-building point of view, it
maybe easier to motivate that at some scale ${\Delta^{\tilde{q}\,LL} }$ and
${\Delta^{\tilde{q}\,RR} }$ are proportional to the unit matrix, meaning that
they possess the full $[U(3)]^3$ flavour symmetry of the gauge sector.
However, below this fundamental scale this symmetry will be reduced to
$[U(2)]^3 \times U(1)$ by renormalisation group (RG) effects from the Yukawa
sector and we restrict our discussion to the $[U(2)]^3 \times U(1)$ case here.
Now we assume that the trilinear terms $A^q_{ij}$ are the spurion fields which
break $[U(2)]^3 \times U(1)$ down to the $U(1)_B$ baryon number symmetry.
(After including the lepton sector the remaining anomaly-free symmetry group
is $U(1)_{B-L}$.)  Note that RG effects do not destroy the symmetry of the
Yukawa sector, because the soft terms do not mix into the Yukawa couplings. By
contrast, in the standard MFV scenario \cite{agis} the role of
flavour-symmetric and symmetry-breaking terms is interchanged, and the latter
(here the Yukawa couplings) mix into the former. Still our scenario is not
completely RG-invariant, because the trilinear terms mix into ${\Delta
  _{ij}^{\tilde{q}\,RR} }$ and ${\Delta _{ij}^{\tilde{q}\,LL} }$.  We next
apply the rotations of \eq{rotq} to the quark supermultiplets of the first two
generations to render the upper left $2\times 2$ submatrix of $\bf A^q$
diagonal and real. That is, the Cabibbo angle arises from the misalignment of
$\bf A^u$ with $\bf A^d$ in flavour space. The $u$, $d$, $s$ and $c$ quark
masses all arise from supersymmetric self-energies involving $A^q_{ii}$. Note
that the RG evolution destroys the $[U(2)]^3 \times U(1)$ symmetry of ${\Delta
  _{ij}^{\tilde{q}\,RR} }$ and ${\Delta _{ij}^{\tilde{q}\,LL} }$ due to
$A^q_{11}\neq A^q_{22}$, but only generates off-diagonal ${\Delta
  _{12}^{\tilde{q}\,LL} }$ terms through tiny electroweak loops as in MFV
scenarios.  The vacuum stability bound in the first row of the table in
Sect.~\ref{sec:comp} is absent and the corresponding bounds on the diagonal
elements $A^q_{ii}$ can only be obtained after loop corrections to the scalar
potential are included. The CKM elements of the third row and column are
obtained by calculating $\Delta U_L^q$ of \eq{DeltaU}.

The idea that flavour violation is a collateral damage of supersymmetry
breaking is not only economical, it also solves one of the most urgent
problems of the MSSM: It was pointed out in \cite{Borzumati:1997bd} that the
phase alignment between $A^q_{ii}$ and the radiatively generated quark
masses suppresses the supersymmetric contribution to the neutron electric
dipole moment. Since $Y^{d_1}=0$, the phase of $\mu$ does not enter the 
neutron EDM at the one-loop level.

\section{Charged-Higgs and chargino couplings}\label{sect:4}

CKM elements do not only enter the Feynman rules for $W$ couplings but also
appear in the couplings of charged Higgs bosons and charginos. The
Feynman rules in the super-CKM scheme i) involve
$V^{(0)}=U_L^{u\,(0)\dag} U_L^{d\,(0)}$ throughout as described in
section 2. Whenever a charged Higgs boson or a chargino couples 
to an external quark there are chirally enhanced one-loop corrections  
similar to those in Figs.~\ref{fig:W} and \ref{fig:dsbW}. 
We can include these diagrams by working with the tree-level
diagrams and replacing $ U_L^{q\,(0)}$ by 
\bea%
U_L^q &=& U_L^{q\,(0)}  \left(1+\Delta U_L^q \right) ,\label{defulq}
\eea%
if the external quark is left-handed.  For instance, we have shown in
Sect.~\ref{sect:2} that the loop corrections to the $\bar{u_f}W^+d_i$ coupling
were correctly included by this replacement (see \eq{physv}). That is, in the
case of $\bar{u_f}W^+d_i$ coupling one simply uses the physical CKM matrix
$V_{fi}$ instead of the tree-level CKM matrix $V^{(0)}_{fi}$. One immediately
notices that (in the super-CKM scheme) the $\widetilde u^*_f W^+\widetilde
d_i$ coupling still involves $V^{(0)}_{fi}$, because the supersymmetric
analogues of the diagrams of Fig.~\ref{fig:W} are not chirally enhanced and
will only lead to small corrections of the typical size of ordinary loop
corrections.  Enhanced corrections to charged-Higgs and
chargino interactions have been discussed for MFV scenarios with large
$\tan\beta$ in Refs.~\cite{Isidori:2001fv, Buras:2002vd}; 
in this section we derive the corresponding results for the non-MFV case 
using the formalism of Sect.~\ref{sect:2}.

Flavour-changing self-energies lead to antihermitian corrections to the
matrices $U^{(0)q}_{L}$. Charged-Higgs and chargino couplings also involve
right-handed fields; the corresponding corrections to $U^{(0)q}_{R}$
are obtained by simply exchanging the chiralities in the expressions for
$\Delta U^q_L$ (cf.\ \eqsand{DeltaU}{defulq}). The CKM matrix which enters
charged-Higgs or chargino vertices is not the physical one, because in these
cases $V^{(0)}$ does not add up to $V$ together with enhanced loop
corrections.  The charged Higgs interaction $\bar{u}_{f} H^+ d_{i}$ has the
Feynman rule 
\begin{equation}
 -i\Lambda^{(0)}_{H^+} = i\left( Y^{u_f \ast} V_{fi}^{(0)} \cos\beta P_L  \; +\;  
V_{fi}^{(0)} Y^{d_i}\sin \beta P_R\right).
\end{equation}
The effect of self-energies in the external legs is included by 
substituting this Feynman rule with 
\begin{eqnarray}
 \Lambda^{(0)}_{H^+}\longrightarrow -\sum\limits_{j,k = 1}^3  && \left[
  \left( 1 + \Delta U_R^{u\dag }  \right)_{fj} 
  Y^{u_j \ast } 
 \left( {1 - \Delta U_L^{u\dag } } \right)_{jk} V_{ki} \cos\beta 
 P_L  \rt. \nn
&& \phantom{\big[} 
\lt.
  \; +\;  
  V_{fj} 
 \left( {1 - \Delta U_L^d } \right)_{jk} Y^{d_k} 
 \left( {1 + \Delta U_R^d } \right)_{ki} \sin\beta P_R  \right].
 \label{higgs1}
\end{eqnarray}
Using 
the explicit expression for $\Delta U_{L,R}^q$ given in \eq{DeltaU}
and expressing  $Y^{q_j}$ in terms of quark masses through \eq{mqdel}  
the substitution rule of \eq{higgs1} becomes
\begin{equation}
\renewcommand{\arraystretch}{1.4}
\begin{array}[b]{l}
   \Lambda^{(0)}_{H^+}  \to  \\[2mm] 
\!
 -\!\sum\limits_{j = 1}^3 \left[ \!\left( \!\!{\begin{array}{c@{\;}c@{\;}c}
   {\frac{{m_{u_1 } }}{{1 + \Delta _{u_1 } }}} & {\frac{{ - \Sigma _{12}^{uRL} }}{{1 + \Delta _{u_2 } }}} & {\frac{{ - \Sigma _{13}^{uRL} }}{{1 + \Delta _{u_3 } }}}  \\
   {\frac{{ - \Sigma _{21}^{uRL} }}{{1 + \Delta _{u_2 } }}} & {\frac{{m_{u_2 } }}{{1 + \Delta _{u_2 } }}} & {\frac{{ - \Sigma _{23}^{uRL} }}{{1 + \Delta _{u_3 } }}}  \\
   {\frac{{ - \Sigma _{31}^{uRL} }}{{1 + \Delta _{u_3 } }}} & {\frac{{ - \Sigma _{32}^{uRL} }}{{1 + \Delta _{u_3 } }}} & {\frac{{m_{u_3 } }}{{1 + \Delta _{u_3 } }}}  \\
 \end{array}} \!\!\!\right)_{\!\!\!\!fj}^{\!\!\! \ds }\!\!
\frac{\ds V_{ji}\cos\beta}{\ds v_u } 
\,P_L  
+ 
\frac{\ds V_{fj}\sin\beta}{\ds v_d}\left( \!\!\begin{array}{c@{\;}c@{\;}c}
    {\frac{{m_{d_1 } }}{{1 + \Delta _{d_1 } }}} & {\frac{{ - \Sigma _{12}^{dLR} }}{{1 + \Delta _{d_2 } }}} & {\frac{{ - \Sigma _{13}^{dLR} }}{{1 + \Delta _{d_3 } }}}  \\
    {\frac{{ - \Sigma _{21}^{dLR} }}{{1 + \Delta _{d_2 } }}} & {\frac{{m_{d_2 } }}{{1 + \Delta _{d_2 } }}} & {\frac{{ - \Sigma _{23}^{dLR} }}{{1 + \Delta _{d_3 } }}}  \\
    {\frac{{ - \Sigma _{31}^{dLR} }}{{1 + \Delta _{d_3 } }}} & {\frac{{ - \Sigma _{32}^{dLR} }}{{1 + \Delta _{d_3 } }}} & {\frac{{m_{d_3 } }}{{1 + \Delta _{d_3 } }}}  \\
\end{array} \!\!\right)_{\!\!ji} \!\!\! P_R \right]  \\ 
 \end{array}
  \label{higgs2}
\end{equation}
We observe a cancellation between the inverse quark masses in $\Delta U_L^q$
(see \eq{DeltaU}) and the factors of $m_{q_i}$ from the $Y^{q_i}$'s in the 
effective off-diagonal couplings. 

For all Higgs processes the genuine vertex correction
$\Lambda^{(1)}_{H^+} $ is of the same order as the diagrams with self
energies in the external leg.  Furthermore, in the absence of terms with
the ``wrong'' vev in the squark mass matrices there is an exact
cancellation between the genuine vertex correction and the external
self-energies in the decoupling limit. This cancellation was observed
for neutral Higgs couplings in Ref.~\cite{cs} and can be understood from
\fig{fig:dec}: The upper right diagram involving $A_{fi}^d$ merely
renormalises the Yukawa coupling and maintains the type-II 2HDM
structure of the tree--level Higgs sector.  Therefore the loop-corrected
Higgs couplings are identical to the tree-level ones, provided they are
expressed in terms of $V_{fi}$ and the physical quark masses. In our
diagrammatic approach $A_{fi}^q$ enters both the proper vertex
correction and $\Sigma _{jk}^{qLR}$ and cancels from the combined
result. 

We neglect all external momenta, so that our expression for
$\Lambda^{(1)}_{H^+}$ is not valid for top or $H^+$ decays unless the
gluino or the squarks appearing in the loop function are much heavier
than the top quark and the charged Higgs boson.  The proper vertex
correction, to be added to \eqsand{higgs1}{higgs2}, reads:
\begin{equation}
\renewcommand{\arraystretch}{1.8}
\begin{array}{l}
\ds  \!\!\!\!\!\Lambda^{(1)}_{H^+} \!=\! - \frac{{2\alpha _s }}{{3\pi }}m_{\tilde g} \sum\limits_{s,t = 1}^6 {\sum\limits_{k,l = 1}^3 {\left\{ {\left( {V_{s\;fk}^{\left( 0 \right)\;u\;LL} V_{t\;li}^{\left( 0 \right)\;d\;RR} P_R  + V_{s\;fk}^{\left( 0 \right)\;u\;RL} V_{t\;li}^{\left( 0 \right)\;d\;RL} P_L } \right)H_{kl}^{ + \,LR} } \right.} }  \\ 
  \,\,\,\;\;\;\;\;\;\;\;\;\;\;\;\;\;\;\;\;\;\;\;\;\;\;\;\;\;\;\;\;\;\;\;\;\;\;\;\;\;\; + \left( {V_{s\;fk}^{\left( 0 \right)\;u\;LR} V_{t\;li}^{\left( 0 \right)\;d\;LR} P_R  + V_{s\;fk}^{\left( 0 \right)\;u\;RR} V_{t\;li}^{\left( 0 \right)\;d\;LL} P_L } \right)H_{kl}^{ + \,RL}  \\ 
  \,\,\,\;\;\;\;\;\;\;\;\;\;\;\;\;\;\;\;\;\;\;\;\;\;\;\;\;\;\;\;\;\;\;\;\;\;\;\;\;\;\; + \left( {V_{s\;fk}^{\left( 0 \right)\;u\;LL} V_{t\;li}^{\left( 0 \right)\;d\;LR} P_R  + V_{s\;fk}^{\left( 0 \right)\;u\;RL} V_{t\;li}^{\left( 0 \right)\;d\;LL} P_L } \right)H_{kl}^{ + \,LL}  \\ 
  \,\left. {\,\;\;\;\;\;\;\;\;\;\;\;\;\;\;\;\;\;\;\;\;\;\;\;\;\;\;\;\;\;\;\;\;\;\;\;\;\;\;\;\, + \left( {V_{s\;fk}^{\left( 0 \right)\;u\;LR} V_{t\;li}^{\left( 0 \right)\;d\;RR} P_R  + V_{s\;fk}^{\left( 0 \right)\;u\;RR} V_{t\;li}^{\left( 0 \right)\;d\;RL} P_L } \right)H_{kl}^{ + \,RR} } \right\} \\ 
  \;\;\;\;\;\;\;\;\;\;\;\;\;\;\;\;\;\;\;\;\;\;\;\;\;\;\;\;\;\;\;\;\;\;\;\;\;\;\;\;\; \times C_0 \left( {m_{\tilde u_s } ,m_{\tilde d_t } ,m_{\tilde g} } \right) \\ 
 \end{array}
 \label{higgsvertex}
 \end{equation}
The coefficients $H_{kl}^{ + \,AB}$ are given in \eq{defhij} of the appendix.

In the case of chargino interactions we must take into account that a squark
never comes with an enhanced self-energy, even if the squark line is an
external line of the considered Feynman diagram.  Here four different
couplings (and their hermitian conjugates) occur.  Neglecting numerical
factors and chargino mixing matrices, the Feynman rules for the chargino
couplings contain the following flavour structures:
\begin{equation}
\renewcommand{\arraystretch}{1.4}
\begin{array}{l@{~~~~~~~~~~~~}l}
\bar{u}_{fL}\, \tilde{\chi}^+ \, \tilde{d}_s: & 
 \sum\limits_{j = 1}^3  {V_{fj}^{(0)} Y^{d_j } W_{j + 3,s}^{\tilde d}P_R
\mbox{~~~~and~~} g_w
 \sum\limits_{j = 1}^3  V_{fj}^{(0)} W_{js}^{\tilde d} }  P_R,  \\
 \bar{u}_{fR}\, \tilde{\chi}^+ \, \tilde{d}_s: & 
 \sum\limits_{j = 1}^3 {Y^{u_f \ast } V_{fj}^{(0)} W_{js}^{\tilde d} } P_L,  
  \\ 
\tilde{u}^*_t \, \tilde{\chi}^+ \, d_{i\,L}: &
 \sum\limits_{j = 1}^3 W_{j + 3,t}^{\tilde u * } Y^{u_j \ast }
 V_{ji}^{(0)} P_L  
\mbox{~~~~and~~} g_w
 \sum\limits_{j = 1}^3  W_{jt}^{\tilde u * } V_{ji}^{(0)}  P_L , \\
\tilde{u}^*_t \, \tilde{\chi}^+ \, d_{i\,R}: &
 \sum\limits_{j = 1}^3 W_{jt}^{\tilde u * } V_{ji}^{(0)} Y^{d_i}  P_R . \\ 
 \end{array}\label{tree-chargino-couplings}
\end{equation}
Here $g_w$ is the SU(2) gauge coupling and $W_{ij}^{\tilde q}$ is defined 
in \eq{defrot}. 
Again we include the self-energy corrections, and express $V^{(0)}$ in terms
of the physical CKM matrix. Then the expressions in
\eq{tree-chargino-couplings} become:

\begin{eqnarray}
\bar{u}_{fL}\, \tilde{\chi}^+ \, \tilde{d}_s: && \sum\limits_{l,m = 1}^3  
V_{fl} 
 \left( {1 - \Delta U_L^d } \right)_{lm} Y^{d_m} W_{m + 3,s}^{\tilde d} 
 P_R \nn
  &&\mbox{and~~~} g_w \sum\limits_{l,m = 1}^3  
  V_{fl} 
    \left( {1 - \Delta U_L^d } \right)_{lm} W_{ms}^{\tilde d}  
   P_R , \no \\[2mm] 
 \bar{u}_{fR}\, \tilde{\chi}^+ \, \tilde{d}_s:&& \sum\limits_{j,k,l,m = 1}^3 
      \left( {1 + \Delta U_L^{u\dag } } \right)_{fj} Y^{u_j \ast } 
      \left( {1 - \Delta U_L^{u\dag } } \right)_{jk} V_{kl} 
      \left( {1 - \Delta U_L^d } \right)_{lm} W_{ms}^{\tilde d}  P_L  
   , \no\\[2mm] 
\tilde{u}^*_t \, \tilde{\chi}^+ \, d_{i\,L}: && 
\sum\limits_{j,k = 1}^3 W_{j + 3,t}^{\tilde u * } Y^{u_j \ast }  
  \left( {1 - \Delta U_L^{u\dag } } \right)_{jk} V_{ki} P_L  
\nn
  &&\mbox{and~~~} g_w
\sum\limits_{j,k = 1}^3  W_{jt}^{\tilde u * } 
  \left( {1 - \Delta U_L^{u\dag } } \right)_{jk} V_{ki} P_L , 
\no\\[2mm] 
\tilde{u}^*_t \, \tilde{\chi}^+ \, d_{i\,R}:&&\sum\limits_{j,k,l,m = 1}^3  W_{jt}^{\tilde u * } 
  \left( {1 - \Delta U_L^{u\dag } } \right)_{jk} V_{kl} 
  \left( {1 - \Delta U_L^d } \right)_{lm} Y^{d_m} 
  \left( {1 + \Delta U_R^d } \right)_{mi} P_R . 
  \label{chargino-couplings}
\end{eqnarray}

We have seen in this section that in the case of non-minimal flavour
violation the CKM matrix (including loop corrections) entering charged Higgs
and quark-squark-chargino vertices is not simply the physical one. Instead it
has to be corrected according to \eq{higgs1} or \eq{higgs2} and
\eq{chargino-couplings}, leading to potentially large effects.

\section{Conclusions}
We have computed the renormalisation of the CKM matrix by
chirally-enhanced flavour-changing SQCD effects in the MSSM with generic
flavour structure.  Our paper extends the work of \cite{Blazek:1995nv},
which considered the MFV case.  We have worked beyond the decoupling
limit $M_{\rm SUSY} \gg v$ and our results are valid for arbitrary
left-right mixing and arbitrary flavour mixing among squarks.
Subsequently we have derived upper bounds on the flavour-changing
off-diagonal elements $\Delta _{ij}^{\tilde{q}\,XY}$ of the squark mass
matrix by requiring that the supersymmetric corrections do not exceed
the measure values of the CKM elements.  For $M_{\rm SUSY}\geq
500\,\gev$ our constraints on \emph{all}\ elements $\Delta
_{ij}^{\tilde{d}\,LR}$, $i<j$, are stronger than the constraints from
FCNC processes. We were further able to derive a strong bound on
$\Delta^{\tilde u\,LR}_{13}$, a quantity which is not constrained by
FCNC's.  For a large value of $\tan\beta$ one can constrain
$\Delta^{\tilde d\,LL}_{13}$ as well.

As an important consequence, we conclude that it is possible to generate the
observed CKM elements completely through finite supersymmetric loop diagrams
\cite{Buchmuller:1982ye,Ferrandis:2004} without violating present-day data on
FCNC processes.  In this scenario the Yukawa sector possesses a higher flavour
symmetry than the trilinear SUSY breaking terms.  Most naturally, first an
exact $[U(2)]^3$ symmetry is imposed on the quark supermultiplets of the first
two generations: Then the corresponding Yukawa couplings $Y^{q}_{ij}$ vanish
and the squark mass terms ${\Delta^{\tilde{q}\,LL}_{ij} }$,
${\Delta^{\tilde{q}\,RR}_{ij} }$ are universal for the first two generations.
In the second step the trilinear terms $A^q_{ij}$ are chosen to break the
flavour symmetry softly and generate light quark masses and off-diagonal CKM
elements radiatively. This result refutes a common conclusion drawn from the
experimental success of the CKM mechanism: It is usually stated that the new
physics of the \tev\ scale must obey the principle of MFV in the sense of
Ref.~\cite{agis}, meaning that the Yukawa couplings are the only spurions
breaking the flavour symmetries. Our analysis has shown that there is a viable
alternative to this scenario: It is well possible that Yukawa couplings obey
an exact flavour symmetry and the spurion fields breaking this symmetry are
the trilinear breaking terms. 

As another application of our results, we have derived supersymmetric loop
corrections to the couplings of charged Higgs bosons and charginos to quarks
and squarks. In these couplings the squark-gluino loops which renormalise the
CKM elements are physical and can have a significant numerical impact because
of their chiral enhancement. We have further pointed out that the calculated
flavour-changing self-energies can have observable effects in the
$W$-mediated production or decay of the top quark, with the SUSY effects 
decoupling as $m_t^2/M_{\rm SUSY}^2$ for $M_{\rm SUSY}\to \infty$.

\section*{Acknowledgements} 
We thank Lars Hofer for sharing the quoted result of Ref.~\cite{hns}
with us.  We thank him and Dominik Scherer for many fruitful
discussions.  We are grateful to Momchil Davidkov for informing us about
mistakes in \eqsand{tree-chargino-couplings}{chargino-couplings} in an
earlier preprint version. A.C.\ is grateful to Martin Gorbahn for
discussions on loop integrals.

This work is supported by BMBF grant 05 HT6VKB and by the EU Contract
No.~MRTN-CT-2006-035482, \lq\lq FLAVIAnet''.

\appendix 

\section*{Appendix: Conventions and Feynman rules}
We denote the tree-level quark mass matrix by ${\bf{m}}_q^{(0)}={\bf Y}^q
v_q$.  The unitary matrices diagonalising these matrices and the squark
mass matrices are denoted by $U_{L,R}^{(0)\,u,d}$ and $W^{\tilde
  u,\tilde d}$, respectively:
\begin{equation}
\renewcommand{\arraystretch}{1.4}
\begin{array}{c}
 U_L^{(0)\,u\dag } {\bf{m}}_u^{(0)} U_R^{(0)\,u}  = 
   {\bf{m}}_u^{\left( D \right)},\qquad\qquad
 U_L^{(0)\,d\dag } {\bf{m}}_d^{(0)} U_R^{(0)\,d}  = 
{\bf{m}}_d^{\left( D \right)}  \\ 
 W^{\tilde u\dag } {\bf M}_{\tilde u}^2 W^{\tilde u}  = 
{\bf M}_{\tilde u}^{2\left( D \right)} , \qquad\qquad 
W^{\tilde d\dag } {\bf M}_{\tilde d}^2 W^{\tilde d}  = 
{\bf M}_{\tilde d}^{2\left( D \right)}  \\ 
 \end{array} \label{defrot}
\end{equation}
The superscript $(D)$ in \eq{defrot} indicates diagonal matrices. 
That is, the mass eigenstates of the quarks and squarks are obtained
from the original fields by unitary rotations in flavour space involving
the matrices $U_L^{(0)\,u,d}$, $U_R^{(0)\,u,d}$ and $W^{\tilde u,\tilde
  d}$ as defined in \cite{Drees}.  The Feynman rules for the
quark-squark-gluino vertices in the basis of mass eigenstates then read
\begin{equation}
 - i\sqrt 2 g_s T^a \sum\limits_{j = 1}^3 {\left( 
   {U_{L\,ji}^{(0)\,q}W_{js}^{\tilde q*} P_L  - 
    U_{R\;ji}^{(0)\,q}W_{j + 3,s}^{\tilde q*} P_R } \right)} \label{qin}
\end{equation}
for an incoming quark and 
\begin{equation}
 - i\sqrt 2 g_s T^a \sum\limits_{j = 1}^3 {\left( 
   {U_{L\,ji}^{(0)\,q*} W_{js}^{\tilde q}P_R  - 
    U_{R\,ji}^{(0)\,q*} W_{j + 3,s}^{\tilde q}P_L } \right)} 
\label{qout}
\end{equation}
for an outgoing quark. The interaction vertices are depicted in
Fig.~\ref{Vertex}.  
\begin{nfigure}{tb}
\centerline{\includegraphics[width=0.7\textwidth]{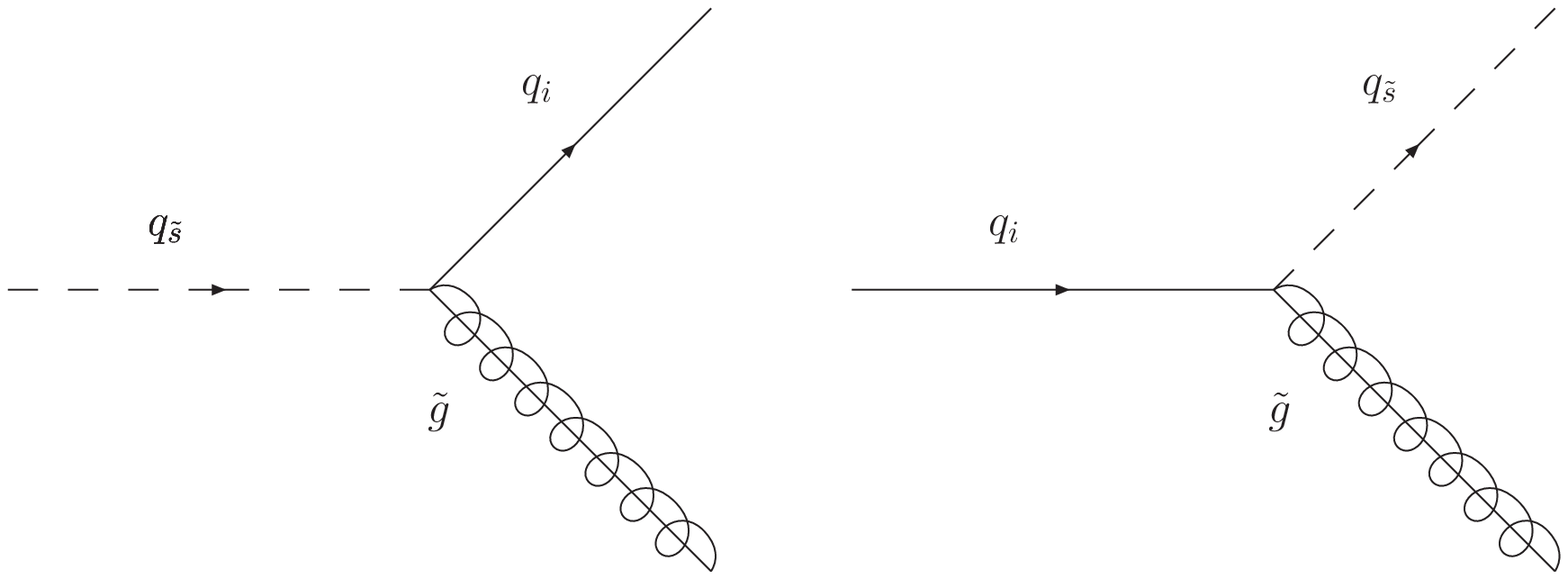}}
\caption{Quark-squark-gluino vertex. The Feynman rules for the left and right
  diagrams  are given in \eqsand{qout}{qin}, respectively.}
\label{Vertex}
\end{nfigure}
\eqsand{qin}{qout} hold in any basis for the quark and squark fields, provided
the quark-squark-gluino coupling is flavour-diagonal in the original basis, in
which the mass matrices $m^{(0)}_{u,d}$ and $M_{\tilde u,d}^2$ are defined.
This condition is not only fulfilled if the original basis consists of 
weak (s)quark eigenstates but also for the super-CKM basis.

Our starting point is a basis of weak eigenstates: 
The squark mass term in the Lagrangian reads 
\begin{eqnarray}
{\cal
  L}_{\tilde m} &=&
-(\tilde d_L^*,\tilde s_L^*,\tilde b_L^*, \tilde d_R^*,
\tilde s_R^*,\tilde b_R^*) {\rm M}_{\tilde d}^{\rm w 2} 
 (\tilde d_L, \tilde s_L,\tilde
b_L, \tilde d_R, \tilde s_R,\tilde b_R)^T \nn
&& - (\tilde u_L^*,\tilde c_L^*,\tilde t_L^*, \tilde u_R^*,
\tilde c_R^*,\tilde t_R^*) {\rm M}_{\tilde u}^{\rm w 2}  
(\tilde u_L, \tilde c_L,\tilde
t_L, \tilde u_R, \tilde c_R,\tilde t_R)^T \no
\end{eqnarray}
with
\begin{equation}
\!\!
\begin{array}{l}
{\rm M}_{\tilde d}^{\rm w 2}   = \left( 
 {\begin{array}{*{20}c}
   \!\!
  {{\bf{M}}_{\tilde q}^2  - 
    {M_Z^2 \left( { 1 + \frac{1}{3}\sin ^2 \theta _W } \right) 
                 \cos 2\beta }  {\bf{1}} + 
   {\bf{m}}_d^{(0)} {\bf{m}}_d^{(0)\dag}  } 
  & 
  {   -v_d {\bf{A}}_w^{d} - \bf{m}_d^{(0)} \mu \tan \beta  }  \\
  {   -v_d {\bf{A}}_w^{d\dag}  - \mu ^* \tan \beta  
      {\bf{m}}_d^{(0)\dag}   }
  &
  \hspace{-7ex} {{\bf{M}}_{\tilde d}^2  - 
    {\frac{1}{3}M_Z^2 \cos 2\beta \sin ^2 \theta _W } {\bf{1}} +
  {\bf{m}}_d^{(0)\dag}  {\bf{m}}_d^{(0)} }  \\
 \end{array}} \right) \\[7mm] 
{\rm M}_{\tilde u}^{\rm w 2}  = \left( 
{\begin{array}{*{20}c}
   \!\!
   {{\bf{M}}_{\tilde q}^2   + 
    {M_Z^2 \left( {1 + \frac{2}{3}\sin ^2 \theta _W } \right)
                 \cos 2\beta } {\bf{1}} + 
   {\bf{m}}_u^{(0)} {\bf{m}}_u^{(0)\dag}  } 
  & 
  {   -v_u {{\bf{A}}_w^{u} - {\bf{m}}_u^{(0)} \mu \cot \beta  }  }\\
   {  -v_u{\bf{A}}_w^u{}^\dag } -  \mu ^* \cot \beta  
       {\bf{m}}_u^{(0)\dag}   
  & 
  \hspace{-7ex} {{\bf{M}}_{\tilde u}^2  + 
   {\frac{2}{3}M_Z^2 \cos 2\beta \sin ^2 \theta _W } {\bf{1}} +
  {\bf m}_u^{(0)^\dag}  {\bf m}_u^{(0)} }  \\
 \end{array}} \right) \\ 
\end{array}\label{defmm}
\end{equation}
The physical CKM matrix differs from 
$V^{(0)} = U_L^{(0)\,u\,\dag} U_L^{(0)\,d}$ 
by the corrections from the finite squark-gluino
self-energies, which are the subject of this paper.  In physical
processes with external quarks the matrices of \eq{defrot} appear in
pairs and it is useful to define:
\begin{equation}
\renewcommand{\arraystretch}{1.4}
\begin{array}{l} 
\ds V^{(0)\,q\, RL}_{s\,fi} \equiv \sum_{j,k=1}^3
 U_{R\, jf}^{(0)\,q*} W_{j + 3,s}^{\tilde q} 
 U_{L\,ki}^{(0)\,q} W_{ks}^{\tilde{q}*} 
\\ 
\ds V^{(0)\,q\, LR}_{s\,fi} \equiv \sum_{j,k=1}^3
U_{L\, jf}^{(0)\,q*} W_{js}^{\tilde q} 
U_{R\,ki}^{(0)\,q} W_{k + 3,s}^{\tilde{q}*} 
\\ 
\ds V^{(0)\,q\, LL}_{s\,fi} \equiv \sum_{j,k=1}^3
 U_{L\,jf}^{(0)\,q*} W_{js}^{\tilde q} U_{L\,ki}^{(0)\,q} W_{ks}^{\tilde{q}*}
\\ 
\ds V^{(0)\,q\, RR}_{s\,fi} \equiv \sum_{j,k=1}^3
 U_{R\,jf}^{(0)\,q*} W_{j + 3,s}^{\tilde q} U_{R\,ki}^{(0)\,q} W_{k+3,s}^{\tilde{q}*}
\\ 
 \end{array}\label{abbr}
\end{equation}
Flavour violation in the squark mass matrices is usually quantified in
the super-CKM basis, in which $m_u^{(0)}=m_u^{\left( D \right)}$ and
$m_d^{(0)}=m_d^{\left( D \right)}$. Then we can use \eq{abbr} and the
Feynman rules of \eqsand{qin}{qout} with the substitutions

\begin{equation}
\sum_{j=1}^3
U_{L\,ji}^{(0)\,q*}W_{js}^{\tilde q}  \to  W_{is}^{\tilde q} , \qquad 
\sum_{j=1}^3 U_{R\,ji}^{(0)\,q*}W_{j + 3,s}^{\tilde q}   \to  
W_{i + 3,s}^{\tilde q} .
\end{equation}
Next we relate the quantities of \eq{massmatrix}, which are defined in the
super-CKM basis, to the parameters of the MSSM Lagrangian, which are defined
in a weak basis. To this end we have to specify a weak basis as our starting
point and we choose a basis with $m_d^{(0)}=m_d^{\left( D \right)}$,
$U_R^{(0)\,u}=U_L^{(0)\,d}=U_R^{(0)\,d}=1$ and $U_L^{(0)\,u}=V^{(0)}$.  Note
that $m_q^{\left( D \right)}$ can be complex, if the threshold corrections
$\Delta_{q_i}$ in \eq{mqdel} are complex.  The transition from this weak basis
to the super-CKM basis only involves a rotation of the left-handed up--type
supermultiplets with $V^{(0)}$. Therefore the down squark mass matrix is
unchanged, ${\rm M}_{\tilde d}^2={\rm M}_{\tilde d}^{\rm w 2}$, while

\begin{eqnarray}
\lefteqn{{\rm M}_{\tilde u}^2  \;=} \label{upsckm}\\
&& \!\!\!\!\!\!\!
\left( 
\begin{array}{*{2}c} 
    \!\! V^{(0)} {\bf{M}}_{\tilde q}^2 V^{(0)\dag}
     + 
    {M_Z^2 \left( {1 + \frac{2}{3}\sin ^2 \theta _W } \right)
                 \cos 2\beta } {\bf{1}} + 
   {\bf m}_u^{(D)} {\bf m}_u^{(D)\, \dag} 
  & 
  \;{ -V^{(0)} {v_u\bf{A}}_w^{u}  - {\bf{m}}_u^{(D)} \mu \cot \beta }   \\
   { -{v_u\bf{A}}_w^u{}^\dag V^{(0)\dag} - \mu ^* \cot \beta } 
       {\bf{m}}_u^{(D)\dag} \quad  
  & 
  \hspace{-12.8ex} {\bf{M}}_{\tilde u}^2  + 
   {\frac{2}{3}M_Z^2 \cos 2\beta \sin ^2 \theta _W } {\bf{1}} +
  {\bf m}_u^{(D)\, \dag} {\bf m}_u^{(D)}\\
 \end{array} \!\!\right) \no
\end{eqnarray}

Thus the trilinear terms of the super-CKM basis, ${\bf{A}}^{q}$,  and 
those in the weak basis are related as

\begin{eqnarray}
{\bf{A}}^{d} &=& {\bf{A}}_w^{d} , \qquad\qquad
{\bf{A}}^{u} \;=\;  V^{(0)\dag} {\bf{A}}_w^{u}, \label{arel}
\end{eqnarray}

and $\Delta _{ij}^{\tilde q \, LR}$ is expressed in terms of 
$A^{q}_{ij}$ in \eq{dea}.
SU(2) symmetry enforces a relation between the upper left $3\times 3$
sub-matrices of ${\rm M}_{\tilde u}^2$ and ${\rm M}_{\tilde d}^2$, since they
both involve ${\bf{M}}_{\tilde q}^2 $. 
The corresponding relation in the super-CKM basis is read off from 
\eq{upsckm}: 
\begin{equation}
\Delta _{ij}^{\tilde d\, LL} = \left[ V^{(0)} {{\rm M}}_{\tilde u}^2 
V^{(0)}\right]_{ij} \label{llrel} 
\end{equation}
for $i,j=1,2,3$ and $i\neq j$. \eq{llrel} was derived in
Ref.~\cite{Misiak:1997ei} with $V^{(0)}=V$, i.e.\ \eq{llrel} generalises the
latter result to the case that radiative corrections to $V$ are included.

The bounds on $\Delta _{ij}^{\tilde q\, XY}$ derived in this paper assume that
the squark-gluino loops at most saturate the measured elements of $V$. The
extremal values for $\Delta _{ij}^{\tilde q\, XY}$ correspond to the case 
$V^{(0)}=1$, for which the super-CKM basis coincides with a weak basis.
This limiting case is realised in the scenario of Sect.~\ref{sect:sof}, 
in which all of $V$ is generated through supersymmetric loops.

In the super-CKM basis the coefficients $H_{ij}^{+\,AB}$ in
\eq{higgsvertex} are given by
\begin{eqnarray}
  H_{ij}^{ + \,LR}  &=& 
    \mu  V_{ij}^{(0)} Y^{d_j } \cos \beta \, -\,  
    \sum\limits_{k = 1}^3 {V_{jk}^{(0)} A_{ki}^d 
                          \sin \beta }  \nn 
  H_{ij}^{ + \,RL}  &=& 
    \mu ^*  V_{ij}^{(0)} Y^{u_i *} \sin \beta  \, -\,  
    \sum\limits_{k = 1}^3 V_{ki}^{(0)} A_{kj}^{u*} 
                           \cos \beta    \nn 
  H_{ij}^{ + \,LL} &=& 
    \sin ( 2 \beta ) \frac{M_W}{\sqrt 2 g_2} \, V_{ij}^{(0)}  
    \left(  |Y^{u_i} |^2  + 
            |Y^{d_j} |^2 
           - g_2^2 \right) 
       \nn 
  H_{ij}^{ + \,RR} &=&  
    \frac{\sqrt 2 M_W}{g_2} 
       Y^{u_i *} V_{ij}^{(0)} Y^{d_j }  \label{defhij}
\end{eqnarray}

Finally we quote our conventions for the two-point, three-point and
four-point one-loop functions $B_0$, $C_0$ and $D_0$:

\begin{eqnarray}
B_0 \left( m_1 ,m_2  \right) &=& 1 + 
\frac{{m_1 ^2 \ln \left( {\frac{{Q ^2 }}{{m_1 ^2 }}} \right) - 
m_2 ^2 \ln \left( {\frac{{Q ^2 }}{{m_2 ^2 }}} \right)}}{m_1 ^2  - m_2 ^2 }
\nn
C_0 \left( {m_1 ,m_2 ,m_3 } \right) &=& 
 \frac{B_0 ( m_1 ,m_2) - B_0 ( m_1 ,m_3) }{m_2^2-m_3^2} \nn
&=&
\frac{{m_1 ^2 m_2 ^2 \ln \left( {\frac{{m_1 ^2 }}{{m_2 ^2 }}} \right) + 
m_2 ^2 m_3 ^2 \ln \left( {\frac{{m_2 ^2 }}{{m_3 ^2 }}} \right) + 
m_3 ^2 m_1 ^2 \ln \left( {\frac{{m_3 ^2 }}{{m_1 ^2 }}} \right)}}{{\left( 
        {m_1 ^2  - m_2 ^2 } \right)\left( {m_2 ^2  - m_3 ^2 } \right)
\left( {m_3 ^2  - m_1 ^2 } \right)}} \nn
D_0 \left( {m_1 ,m_2 ,m_3,m_4 } \right) &=& 
 \frac{C_0 ( m_1 ,m_2,m_3) - C_0 ( m_1 ,m_2,m_4) }{m_3^2-m_4^2} \no
\end{eqnarray}
The two-point function $B_0$ is UV-divergent, our definition above 
is $\ov{\rm MS}$-subtracted. UV divergence and the renormalisation scale 
$Q$ drop out from our results thanks to the super-GIM mechanism.

\end{document}